\documentclass[aip, jcp, amsmath,amssymb, reprint, floatfix,citeautoscript]{revtex4-1}

\usepackage{graphicx}
\usepackage{dcolumn}
\usepackage{bm}

\usepackage{braket}
\usepackage[usenames,dvipsnames]{color}

\usepackage{ragged2e}
\usepackage[labelsep=period, format=plain, justification=justified, font=footnotesize]{caption}
\usepackage{times}
\usepackage{txfonts}
\usepackage{relsize}
\usepackage{multirow}
\usepackage{tabularx,booktabs}
\newcolumntype{Y}{>{\centering\arraybackslash}X}
\usepackage{threeparttable}
\usepackage[version=3]{mhchem}
\usepackage{mathtools}

\DeclareCaptionJustification{myjust}{\justifying}
\newcommand{\RNum}[1]{\uppercase\expandafter{\romannumeral #1\relax}}

\begin{document}

\title{Excited state diabatization on the cheap using DFT: Photoinduced electron and hole transfer}

\author{Yuezhi Mao}
\author{Andr\'{e}s Montoya-Castillo}
\author{Thomas E. Markland}
\email{tmarkland@stanford.edu}
\affiliation{Department of Chemistry, Stanford University, Stanford, California, 94305, USA}

\date{\today}

\begin{abstract}
	Excited state electron and hole transfer underpin fundamental steps in processes such as exciton dissociation at photovoltaic heterojunctions, photoinduced charge transfer at electrodes, and electron transfer in photosynthetic reaction centers. Diabatic states corresponding to charge or excitation localized species, such as locally excited and charge transfer states, provide a physically intuitive framework to simulate and understand these processes. However, obtaining accurate diabatic states and their couplings from adiabatic electronic states generally leads to inaccurate results when combined with low-tier electronic structure methods, such as time dependent density functional theory (TDDFT), and exorbitant computational cost when combined with high-level wavefunction-based methods. Here we introduce a DFT-based diabatization scheme, $\Delta$-ALMO(MSDFT2), which directly constructs the diabatic states using absolutely localized molecular orbitals (ALMOs). We demonstrate that our method, which combines ALMO calculations with the $\Delta$SCF technique to construct electronically excited diabatic states and obtains their couplings with charge-transfer states using our MSDFT2 scheme, gives accurate results for excited state electron and hole transfer in both charged and uncharged systems that underlie DNA repair, charge separation in donor-acceptor dyads, chromophore-to-solvent electron transfer, and singlet fission. This framework for the accurate and efficient construction of excited state diabats and evaluation of their couplings directly from DFT thus offers a route to simulate and elucidate photoinduced electron and hole transfer in large disordered systems, such as those encountered in the condensed phase.
\end{abstract}

{\maketitle}
\normalsize

\section{Introduction}

Photoinduced electron transfer (ET) is a fundamental step in important photochemical processes ranging from carrier generation and transport in photovoltaic materials such as organic solar cells and dye-semiconductor heterostructures, to photo-damage of biomolecules such as DNA and RNA, and photo-initiated catalytic processes including reduction of \ce{CO2} by photoactive transition metal complexes, water oxidation at photoelectrodes, and hot electron transfer from plasmonic nanoparticles to molecules. Charge-localized diabatic states (diabats) provide a chemically intuitive basis to simulate and elucidate these photoinduced processes since they possess the desirable feature that their chemical character is robust to changes in nuclear configurations. Indeed, diabats form the basis of widely used theories of charge transfer, such as Marcus-Hush theory \cite{Marcus1956, Hush1961, Marcus1993}, where diabatic state energies and couplings are essential to determine reaction rates. These diabatic states, especially when it is feasible to compute their associated nuclear forces, also permit a fully dynamical description of excited state charge-transfer processes that provides mechanistic details beyond those accessible from rate theories. Accurate and efficient ways to construct such diabats and calculate their couplings therefore open the door to studying the nonadiabatic dynamics of out-of-equilibrium photoinduced processes occurring across multiple time and length scales.

In photoinduced electron transfer processes, the reactant state typically features a photo-excitation on the donor molecule while the product state features electron-hole separation. The former can be physically represented by a locally excited (LE) diabat and the latter a charge-transfer (CT) diabat. Density functional theory (DFT) \cite{Hohenberg1964,Kohn1965} offers an appealing approach to constructing these diabats due to its computationally efficient treatment of electron correlation. One commonly used approach that can be applied to both wavefunction-based theories and DFT is to compute a set of adiabatic states and then transform them into diabats through a unitary rotation. This adiabatic-to-diabatic (ATD) transformation can be constructed by defining diabats as eigenstates of operators that are closely related to the ET process \cite{Cave1996,Cave1997,Voityuk2002,Voityuk2013,Hsu2008,Hsu2009,You2010} or states that are maximally localized on the donor or acceptor moieties \cite{Subotnik2008,Subotnik2009,Subotnik2010,Subotnik2015}. The most commonly employed approach to generate excited states from DFT is through linear-response time-dependent density functional theory (LR-TDDFT)\cite{Runge1984,Dreuw2005,Casida2012}. However, due to self-interaction error, LR-TDDFT underestimates the energy of charger-transfer excited states\cite{Dreuw2004,Dreuw2005} that are essential to photoinduced ET processes. Although these issues can be somewhat alleviated by range-separated hybrid functionals \cite{Tawada2004,Yanai2004,Chai2008b,Rohrdanz2009} and system-specific $\omega$-tuning techniques,\cite{Stein2009,Baer2010} these often require balancing improving the accuracy of CT-type excitation energies with artificial overestimation of LE state energies. In addition, while the application of ATD diabatization to ground-state electron or hole transfer (ET/HT) typically only requires computing the ground and the first excited adiabatic states, in photoinduced ET processes many higher-energy excited states of distinct characters can contribute to the relevant diabats. Since ATD-based approaches, such as the generalized Mulliken-Hush (GMH) method,\cite{Cave1996,Cave1997} can usually only be applied to a few adiabatic states, these must be carefully selected through a comprehensive excited-state analysis to achieve reliable results. In addition, diabatic states constructed through an ATD procedure are typically not variationally optimized and the transformation matrix itself varies with the nuclear positions, which complicates access to forces needed for geometry optimization and nuclear dynamics on diabatic potential energy surfaces.

These drawbacks of ATD-based diabatization schemes have motivated the development of DFT-based approaches that create diabatic states without first computing the excited-state adiabats. However, whereas for ground-state ET and HT there are many DFT-based methods to directly construct diabats and evaluate their couplings, \cite{Kondov2007,Senthilkumar2003,Oberhofer2012,Schober2016,Wu2006,VanVoorhis2010,Pavanello2013,Ramos2014,Cembran2009,Ren2016,Mao2019} fewer approaches exist to generate the LE and CT states and their couplings needed to describe photoinduced ET. One such method \cite{Difley2011} constructs LE states from an LR-TDDFT calculation and the CT states using constrained DFT (CDFT) \cite{Wu2005, Wu2006b} and evaluates LE-CT couplings using the CDFT configuration interaction (CDFT-CI) prescription \cite{Wu2006, Wu2007}. However, its applicability is limited by its strict requirement that the global system TDDFT excited state be well-localized on the donor moiety such that it can be directly treated as the LE state. To enforce the locality of the LE diabat, one can perform an LR-TDDFT calculation on the donor fragment in isolation and then construct the global system diabat by forming an antisymmetrized product with the ground state wavefunctions of the other fragments. A recent work \cite{Chan2013} used this procedure followed by a Thouless transformation \cite{Thouless1960, Difley2011} to construct the LE state and then constructs the CT diabat using the block-localized Kohn-Sham (BLKS) \cite{Mo2007, Cembran2009} approach. However, the use of two different theories (LR-TDDFT and BLKS) for the construction of the LE and CT states means that these are not constructed on an equal footing and the use of LR-TDDFT for the LE state means that this state is not variationally relaxed.

Here we present a new approach to construct variationally optimized LE and CT diabats using absolutely localized molecular orbitals (ALMOs) and demonstrate its accuracy in treating photoinduced electron and hole transfer in model complexes inspired by excited-state processes in biochemical and photovoltaic applications. Our approach exploits $\Delta$SCF\cite{Ziegler1977,Kowalczyk2011b} to target molecular excited states while maintaining balanced accuracy for different types of electronic excited states \cite{Kowalczyk2011b,Barca2018} and the ability of ALMOs\cite{Khaliullin2006, Mao2019} to produce diabatic states that are variationally optimized at the supersystem KS-DFT level. We then couple these diabatic states using a generalization of our ALMO(MSDFT2) scheme, which we have previously demonstrated to give errors below 5\% over a broad range of ground state electron/hole transfer chemical systems\cite{Mao2019}. The resulting approach circumvents the trade-off between the accuracy of the LE and CT states in TDDFT and the use of variationally optimized ALMOs allows one to easily obtain gradients with respect to nuclear configurations that are not easily accessible using ATD-type diabatization methods. We then show how our new method performs in comparison to benchmark values obtained from excited state calculations at the equation-of-motion coupled-cluster singles and doubles (EOM-CCSD) \cite{Stanton1994, Krylov2008} level of theory followed by GMH diabatization\cite{Cave1996, Cave1997} on a variety of systems relevant to DNA photo-damage/protection, charge separation in photovoltaics, photoinduced electron transfer from chromophore to solvent, and singlet fission. By doing this we show that our approach provides an accurate and efficient method for constructing DFT-based diabatic states and evaluating their couplings for photoinduced electron and hole transfer that can easily provide access to nuclear forces for quantum dynamics simulations.

\section{Methods} \label{sec:method}

\subsection{LE and CT states from ALMO calculations} \label{subsec:almo_diabats}

To provide a DFT-based method to create the electronic states that govern photoinduced electron transfer processes, here we show how to construct locally excited (LE) and charge-transfer (CT) diabats and compute their couplings using ALMOs. We have previously shown\cite{Mao2019} for electron and hole transfer in systems in their ground electronic states that using ALMOs prevents charge delocalization between donor and acceptor moieties under the Mulliken definition of charge population,\cite{Mulliken1955} making it suitable for the construction of charge-localized diabats. Here, we consider photoinduced electron transfer in a donor-acceptor complex $D\cdots A$, where the donor ($D$) has $n_D$ electrons and the acceptor ($A$) has $n_A$ electrons. The ground state of this system, $\ket{DA}$, can be represented as
\begin{equation}
\ket{DA} = \mathcal{N} \mathrm{det} \left\lbrace \phi_{D1}, \dots, \phi_{Di} \dots, \phi_{Dn_{D}},  \phi_{A1},  \dots, \phi_{An_{A}} \right \rbrace,
\label{eq:almo_gs}
\end{equation}
where ``det'' denotes the Slater determinant and $\mathcal{N} = 1/\sqrt{(n_D + n_A)!}$ is the normalization constant. Each molecular orbital (MO) in the determinant is assigned to either the donor or the acceptor fragment and is expanded using atomic orbital (AO) basis functions located only on that fragment, yielding orbitals that are ``absolutely localized''.

For systems in their ground electronic states that can be readily partitioned into donor and acceptor moieties, one can use a ``bottom-up'' approach to construct the ALMO diabats.~\cite{Mao2019} This consists of performing SCF calculations for the isolated fragments $D$ and $A$ first and then generating the initial guess to the ALMO state by assembling the resulting fragment orbitals via Eq.~(\ref{eq:almo_gs}). This state is also known as the frozen state in ALMO-based energy decomposition analysis.\cite{Khaliullin2007, Horn2016c} One can then variationally optimize the ALMOs by minimizing the Kohn-Sham (KS) energy functional of the full system, $E^{\text{KS}}[\mathbf{P}]$, with respect to the occupied-virtual mixings on each fragment, where $\mathbf{P}$ is the one-particle density matrix (1PDM) that can be constructed from the occupied ALMOs ($\mathbf{C}_{\text{o}}$) through
\begin{equation}
\mathbf{P} = \mathbf{C}_{\text{o}} (\boldsymbol{\sigma}_{\text{oo}})^{-1} \mathbf{C}_{\text{o}}, \quad \boldsymbol{\sigma}_{\text{oo}} = \mathbf{C}_{\text{o}}^{T} \mathbf{S} \mathbf{C}_{\text{o}}.
\label{eq:1pdm}
\end{equation}
Here $\boldsymbol{\sigma}_{\text{oo}}$ is the overlap metric of occupied orbitals, which is obtained by transforming the AO overlap matrix ($\mathbf{S}$) into the ALMO basis.

To create diabats with the donor moiety in its electronically excited state, one can utilize the $\Delta$SCF approach \cite{Ziegler1977,Kowalczyk2011b} within ALMO-based DFT calculations. The $\Delta$SCF approach has been shown to accurately capture both valence and charge-transfer excited states in cases where the target excited state is dominated by the transition between a specific pair of orbitals. \cite{Kowalczyk2011b,Barca2018} This procedure is typically applied by taking the electronic ground state orbitals and then promoting electrons from the occupied to the virtual MOs to generate a non-aufbau electronic configuration. The MOs are then relaxed within this non-aufbau electronic configuration, resulting in an SCF solution that corresponds to an electronic excited state. To prevent the variational optimization from collapsing to the ground state, one can employ the maximum overlap method (MOM)\cite{Gilbert2008} or its initial MOM (IMOM) \cite{Barca2018} variant. In these methods, the occupied orbitals at each SCF iteration are selected to be the ones that have the largest overlap with the span of the occupied orbitals in the previous (MOM) or initial (IMOM) non-aufbau electronic configuration. Hence they are compatible with SCF algorithms performing diagonalization of the Fock matrix, such as the widely used direct inversion of iterative subspace (DIIS) \cite{Pulay1982} method.

\begin{figure*}[t!]
	\centering
	\includegraphics[width=0.75\textwidth]{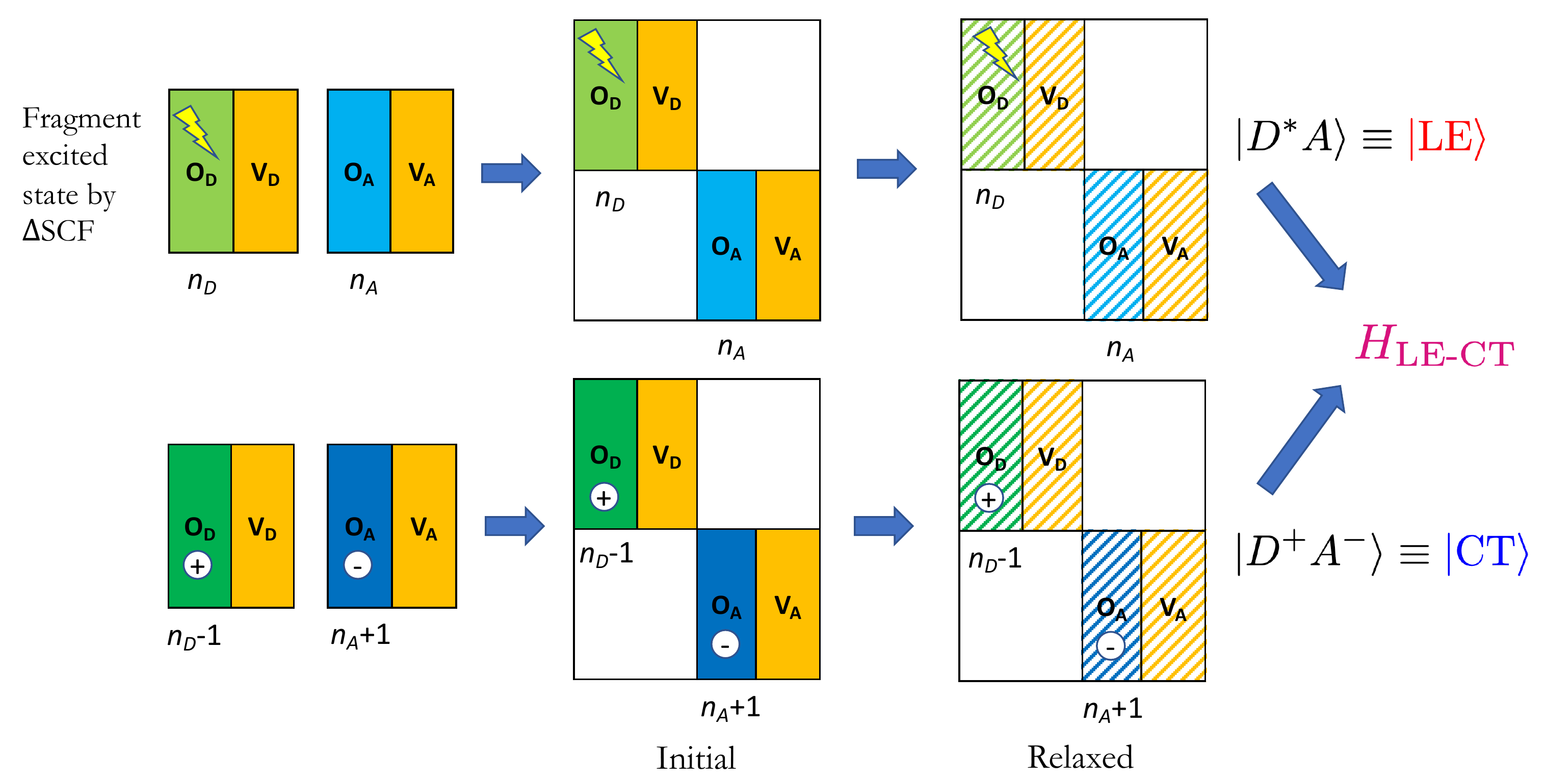}
	\caption{Procedure of our ALMO-based approach for the calculation of LE-CT diabatic couplings. Top row: construction of the LE diabat $\ket{D^\ast A}$ using the ALMO-$\Delta$SCF method; bottom row: construction of the CT diabat $\ket{D^+ A^-}$ using a ground state ALMO calculation. The solid blocks represent unrelaxed fragment orbitals and the shaded blocks represent orbitals variationally relaxed in the full system.}
	\label{fig:almo_msdft_scheme}
\end{figure*}

To generate a variationally optimized LE state from the fragment orbitals of the excited-state donor and ground-state acceptor, one brings together these orbitals and then relaxes them via an additional $\Delta$SCF-type procedure, yielding a global system ALMO state that corresponds to the mutually relaxed LE diabat $\ket{D^\ast A}$ (illustrated in the top row of Fig.~\ref{fig:almo_msdft_scheme}). One begins by calculating the ground state of the acceptor using standard SCF and the excited state of the donor using $\Delta$SCF and then concatenates the fragment orbitals obtained into the form of a single Slater determinant given by
\begin{equation}
\ket{D^\ast A}_{0} = \mathcal{N} \mathrm{det} \left\lbrace \phi^{(0)}_{D1}, \dots, \phi^{(0)}_{Da} \dots, \phi^{(0)}_{Dn_{D}},  \phi^{(0)}_{A1},  \dots, \phi^{(0)}_{An_{A}} \right \rbrace,
\label{eq:almo_le_init}
\end{equation}
which we denote as $\ket{D^\ast A}_0$ and corresponds to the ``Initial'' unrelaxed state on the top row of Fig.~\ref{fig:almo_msdft_scheme}. This initial LE state corresponds to an excitation on the donor dominated by the transition from its $i$-th occupied orbital to $a$-th virtual orbital.

To variationally relax the ALMOs in the LE state we then solve the locally projected SCF (Stoll) equation \cite{Stoll1980, Khaliullin2006} for each fragment,
\begin{equation}
[(\mathbf{I} - \mathbf{SP} + \mathbf{SP}_{X}^{T}) \mathbf{F} (\mathbf{I} - \mathbf{PS} + \mathbf{P}_{X}\mathbf{S})]_{XX} \mathbf{C}_{X} =  \mathbf{S}_{XX} \mathbf{C}_{X} \boldsymbol{\epsilon}_{X}.
\label{eq:stoll_eqn}
\end{equation}
Here $X \in \{D, A \}$ is the fragment index, $\mathbf{F} \equiv \mathbf{F}[\mathbf{P}]$ is the AO Fock matrix constructed from the 1PDM ($\mathbf{P}$), $\mathbf{P}_{X}$ is the projector for fragment $X$, and $\boldsymbol{\epsilon}_{X}$ and $\mathbf{C}_X$ are the diagonal matrix of eigenvalues and the matrix composed of eigenvectors which correspond to the orbital energies and AO expansion coefficients of ALMOs on fragment $X$, respectively. $\boldsymbol{\epsilon}_{X}$ and $\mathbf{C}_X$ are obtained by diagonalizing the locally projected Fock matrix, defined as the quantity in the square bracket on the left-hand side of Eq.~(\ref{eq:stoll_eqn}). In Stoll's equation, the projector for fragment $X$ has the form
\begin{equation}
\mathbf{P}_X = [\mathbf{C}_{\text{o}} (\boldsymbol{\sigma}_{\text{oo}}^{-1})]_{X} \mathbf{C}_{\text{o}, X}^{T}.
\end{equation}
Here the subscript $X$ denotes matrix columns that belong to the partition corresponding to fragment $X$. Explicitly, when $X = D$, one takes columns $1$ to $n_D$ of the resulting matrix and when $X= A$, one takes columns $n_D + 1$ to $n_D + n_A$. The matrices $[\mathbf{C}_{\text{o}} (\boldsymbol{\sigma}_{\text{oo}}^{-1})]_{X}$ and $\mathbf{C}_{\text{o},X}$ are thus the contravariant and covariant versions of the occupied ALMO coefficients belonging to fragment $X$, respectively.\cite{Khaliullin2006,Horn2013} The outer product of $[\mathbf{C}_{\text{o}} (\boldsymbol{\sigma}_{\text{oo}}^{-1})]_{X}$ and $\mathbf{C}_{\text{o},X}$ then gives the Stoll projector to fragment $X$.\cite{Stoll1980, Khaliullin2006} While iterating the Stoll equation (Eq.~(\ref{eq:stoll_eqn})) to self-consistency with only fragments in their ground states yields the relaxed ALMOs, when any of the fragments is in its excited state, this procedure alone would lead to the collapse of this local excited state to the ground state. Hence the Stoll iteration must be performed in concert with a procedure to prevent this excited state collapse.

Here we employ the IMOM procedure\cite{Barca2018} to prevent the excited state from collapsing to the ground state when solving the Stoll equation for the excited fragment $D^\ast$. This procedure requires one to calculate the overlap ($\mathbf{O}$) between ALMOs on fragment $D^\ast$ obtained at a given iteration and the initial occupied MOs on the same fragment, denoted as $\mathbf{C}_{D^\ast,\text{o}}^{(0)}$,
\begin{equation}
\mathbf{O} = (\mathbf{C}_{D^\ast,\text{o}}^{(0)})^{T} \mathbf{S} \mathbf{C}_{D^\ast}.
\end{equation}
To select the orbitals to be occupied for the relaxed fragments, one has to choose the orbitals that have the largest overlap with the initial set of occupied orbitals in the non-aufbau electronic configuration. These orbitals are selected as the $n_D$ orbitals that possess the largest projection onto the span of the initial set of occupied orbitals, $P_{D^\ast j}$, 
\begin{equation}
P_{D^\ast j} = \left(\sum_{i=1}^{n_D} O_{D^\ast i D^\ast j}^2 \right)^{1/2},
\end{equation}
where $i$ and $j$ denote the molecular orbitals in the initial and current iterations, respectively. The selected orbitals are then employed to generate the updated 1PDM. Iterating Eq.~(\ref{eq:stoll_eqn}) with IMOM to self-consistency, one obtains the relaxed LE state, denoted as the ``Relaxed'' state on the top row of Fig.~\ref{fig:almo_msdft_scheme},
\begin{equation}
\ket{D^\ast A} = \mathcal{N} \mathrm{det} \left\lbrace \phi^{(\text{LE})}_{D1}, \dots, \phi^{(\text{LE})}_{Da} \dots, \phi^{(\text{LE})}_{Dn_{D}},  \phi^{(\text{LE})}_{A1},  \dots, \phi^{(\text{LE})}_{An_{A}} \right \rbrace,
\label{eq:almo_le_relaxed}
\end{equation}
where the superscript (LE) indicates that the ALMOs are optimized within the LE state in contrast to the unrelaxed fragment orbitals in $\ket{D^\ast A}_0$. We denote the procedure to variationally optimize ALMOs in the excited state introduced in the previous paragraphs as ALMO-$\Delta$SCF in the following discussion.

To prepare the lowest-energy intermolecular CT diabat we employ the procedure introduced in our previous work.\cite{Mao2019} This CT diabat usually corresponds to transferring an electron from the highest occupied molecular orbital (HOMO) of the donor to the lowest unoccupied molecular orbital (LUMO) of the acceptor, which we denote as $\ket{D^{+}A^{-}}$. One can prepare this diabat by performing an unrestricted ALMO calculation in this charge-separated state.\cite{Horn2013} Starting from SCF calculations for fragments $D^{+}$ and $A^{-}$ in isolation, we construct the CT diabat and then relax its orbitals using the standard ALMO-based SCF procedure, yielding
\begin{equation}
\ket{D^{+}A^{-}} = \mathcal{N} \mathrm{det} \left\lbrace \phi^{(\text{CT})}_{D1},  \dots, \phi^{(\text{CT})}_{Dn_{D}-1},  \phi^{(\text{CT})}_{A1},  \dots, \phi^{(\text{CT})}_{An_{A}}, \phi^{(\text{CT})}_{An_{A}+1} \right \rbrace.
\label{eq:almo_ct_relaxed}
\end{equation}
This procedure is illustrated on the bottom row of Fig.~\ref{fig:almo_msdft_scheme}. As in Eq.~(\ref{eq:almo_le_relaxed}), the superscript (CT) indicates that the ALMOs are relaxed within the specific electronic configuration of the CT state. While the diabat given in Eq.~(\ref{eq:almo_ct_relaxed}) corresponds to an aufbau CT configuration, one could generate higher-energy CT states by employing the ALMO-$\Delta$SCF procedure used to prepare the LE states described above.

\subsection{Diabatic coupling between ALMO LE and CT states} \label{subsec:almo_coupling}

To evaluate the electronic coupling between LE and CT diabats, $\braket{D^\ast A | \hat{H} | D^{+} A^{-}}$, one can utilize the original multistate DFT (MSDFT) approach \cite{Cembran2009, Ren2016} or the MSDFT2 scheme that we proposed in our previous work.\cite{Mao2019} Defining $\ket{\psi_a} \equiv \ket{D^\ast A}$ and $\quad \ket{\psi_b} \equiv \ket{D^+ A^-}$ one can construct the diabatic Hamiltonian in the non-orthogonal ALMO basis
\begin{equation}
\mathbf{H}^{\prime} = 
\begin{pmatrix}
H^{\prime}_{aa} & H^{\prime}_{ab} \\
H^{\prime}_{ba} & H^{\prime}_{bb}
\end{pmatrix}.
\end{equation}
To extract the electronic coupling, $H_{ab}=\braket{D^\ast A | \hat{H} | D^{+} A^{-}}$, between these two ALMO diabats, one transforms $\mathbf{H}^{\prime}$ into an orthogonal basis. Using L\"owdin's symmetric orthogonalization scheme,\cite{Lowdin1950} one obtains $\mathbf{H} = \boldsymbol{\mathcal{S}}^{-1/2} \mathbf{H}^{\prime} \boldsymbol{\mathcal{S}}^{-1/2}$ where the interstate overlap matrix is $[\boldsymbol{\mathcal{S}}]_{xy} \equiv \braket{\psi_x|\psi_y}$ and $x, y \in \{a, b\}$. In the 2-state case this yields
\begin{equation}
H_{ab} = \frac{1}{1- \mathcal{S}_{ab}^2} \left\vert H^{\prime}_{ab} - \frac{H^{\prime}_{aa} + H^{\prime}_{bb}}{2} \mathcal{S}_{ab} \right\vert,
\label{eq:msdft_2state}
\end{equation}
where $\mathcal{S}_{ab}$ is given by
\begin{equation}
\mathcal{S}_{ab} \equiv \Braket{\psi_a | \psi_b} = \mathrm{det}[(\mathbf{C}^{(a)}_{\text{o}})^{T} \mathbf{S} \mathbf{C}^{(b)}_{\text{o}}].
\label{eq:state_overlap}
\end{equation}
Here $\mathbf{C}^{(a)}_{\text{o}}$ and $\mathbf{C}^{(b)}_{\text{o}}$ are ALMO coefficient matrices associated with the two diabats and $\mathbf{S}$ is the AO overlap matrix. In the following discussion, we use $\mathbf{S}_{\text{oo}}^{(ab)} \equiv (\mathbf{C}^{(a)}_{\text{o}})^{T} \mathbf{S} \mathbf{C}^{(b)}_{\text{o}}$ to denote the overlap matrix between occupied orbitals from diabats $\ket{\psi_{a}}$ and $\ket{\psi_{b}}$.

In both MSDFT \cite{Cembran2009, Ren2016} and MSDFT2 \cite{Mao2019} approaches, the diagonal elements of $\mathbf{H}^{\prime}$ are the KS energies of each diabat, i.e., $H^{\prime}_{aa} = E_{a}^{\mathrm{KS}}[\mathbf{P}^{(a)}]$ and $H^{\prime}_{bb} = E_{b}^{\mathrm{KS}}[\mathbf{P}^{(b)}]$, where $\mathbf{P}^{(a)}$ and $\mathbf{P}^{(b)}$ are the 1PDMs associated with $\ket{\psi_a}$ and $\ket{\psi_b}$, respectively. These two schemes differ in the way they approximate the off-diagonal element $H_{ab}^{\prime}$. In the original MSDFT scheme, $H_{ab}^{\prime}$ is given by a KS-DFT correction to the Hartree-Fock (HF) interstate coupling \cite{Amos1961, Thom2009} result:
\begin{equation}
H^{\prime}_{ab} = S_{ab} \left[V_{\text{nn}} + \mathbf{P}_{ab}\cdot\mathbf{h} + \frac{1}{2} \mathbf{P}_{ab} \cdot \mathbf{\RNum{2}} \cdot \mathbf{P}_{ab} + \frac{1}{2}(\Delta E_{a}^{\text{c}} + \Delta E_{b}^{\text{c}})  \right],
\label{eq:msdft1}
\end{equation}
where the first three terms on the right-hand side arise from the nuclear repulsion energy ($V_{\text{nn}}$), one-electron Hamiltonian ($\mathbf{h}$), and electron-repulsion integrals ($\mathbf{\RNum{2}}$) as in the HF theory. $\mathbf{P}_{ab}$ is the transition density matrix between diabats $\ket{\psi_a}$ and $\ket{\psi_b}$
\begin{equation}
\mathbf{P}_{ab} = \mathbf{C}_{\text{o}}^{(a)} \left[ \mathbf{S}_{\text{oo}}^{(ba)} \right]^{-1} (\mathbf{C}_{\text{o}}^{(b)})^{T}.
\label{eq:trans_den}
\end{equation}
The last term in Eq.~(\ref{eq:msdft1}) is a correction intended to incorporate the contribution from the exchange-correlation (XC) functional to the interstate coupling, which is given by the average of the difference between the KS and HF energies of each diabat calculated using their individual 1PDMs
\begin{align}
\Delta E_{a}^{\mathrm{c}} &= E^{\mathrm{KS}}_{a} [\mathbf{P}^{(a)}] - E^{\mathrm{HF}}_{a} [\mathbf{P}^{(a)}], \\
\Delta E_{b}^{\mathrm{c}} &= E^{\mathrm{KS}}_{b} [\mathbf{P}^{(b)}] - E^{\mathrm{HF}}_{b} [\mathbf{P}^{(b)}]. \label{eq:msdft1_xc}
\end{align}

As shown in our previous work,\cite{Mao2019} the treatment of interstate coupling in the original MSDFT approach (Eq.~(\ref{eq:msdft1})) effectively couples the KS determinants using only the HF Hamiltonian once substituted into Eq.~(\ref{eq:msdft_2state}), and is thus equivalent to how the interstate couplings are evaluated in some recently introduced ab initio exciton models \cite{Morrison2014, Morrison2015, Fujita2018}. Our MSDFT2 approach explicitly incorporates the XC contribution in the interstate coupling by utilizing the KS energy functional of the symmetrized transition density matrix ($\tilde{\mathbf{P}}_{ab}$):\cite{Mao2019}
\begin{equation}
H^{\prime}_{ab} = S_{ab} \left[V_{\text{nn}} + \mathbf{P}_{ab}\cdot\mathbf{h} + \frac{1}{2} \mathbf{P}_{ab} \cdot \mathbf{\RNum{2}} \cdot \mathbf{P}_{ab} + E_{\text{xc}}[\tilde{\mathbf{P}}_{ab}]  \right],
\label{eq:msdft2}
\end{equation}
where $\tilde{\mathbf{P}}_{ab} = (\mathbf{P}_{ab} + \mathbf{P}_{ba})/2$. Importantly, the electron-repulsion integrals, $\mathbf{\RNum{2}}$, in Eq.~(\ref{eq:msdft2}) account for Coulomb integrals as well as a fraction of HF exchange when hybrid functionals are used, which differs from that in Eq.~(\ref{eq:msdft1}) where $\mathbf{\RNum{2}}$ denotes the full HF Coulomb and exchange integrals.

A notable feature of the MSDFT2 approach is that the off-diagonal element of the diabatic Hamiltonian given by Eq.~(\ref{eq:msdft2}) reduces to the KS energy of a diabat when $a = b$, ensuring internal consistency of the theory. The same formula (Eq.~(\ref{eq:msdft2})) was also employed to evaluate the coupling between DFT-based diabats in the frozen density embedding (FDE) method for ground-state ET/HT. \cite{Pavanello2013, Ramos2014, Ramos2015} We have previously shown \cite{Mao2019} that for ground-state ET and HT processes, the ALMO(MSDFT2) method can accurately predict diabatic couplings between donor and acceptor states and that its performance is superior to the original MSDFT approach and other popular DFT-based diabatization schemes \cite{Kondov2007,Senthilkumar2003,Oberhofer2012,Schober2016,Wu2006,VanVoorhis2010}. In this work, we assess the ability of both MSDFT2 and MSDFT in coupling LE diabats constructed from ALMO-$\Delta$SCF calculations with ALMO CT states. The overall schemes are denoted as $\Delta$-ALMO(MSDFT2) and $\Delta$-ALMO(MSDFT) in the following discussion.

\subsection{MSDFT2 in the weak-coupling regime} \label{subsec:fock_approach}

In the following, we discuss the weak-coupling regime where the diabatic coupling between two diabats approaches zero, which presents a challenge when calculating the diabatic coupling using ALMO(MSDFT2). We then provide a way to determine when one is in this regime and a physically intuitive alternative coupling scheme that yields stable results in this limit. Two situations that lead to weak coupling are when the overlap between two diabats approaches zero due to increasing intermolecular distance and when there is destructive interference between the orbitals involved in the charge transfer process.

The former situation, which leads to near-zero overlap between the orbitals relevant to the charge transfer process, precludes the construction of the transition density matrix between two ALMO diabats. In particular, the matrix inversion of the overlap matrix between the diabats' occupied orbitals ($\mathbf{S}_{\text{oo}}^{(ba)}$) required in Eq.~(\ref{eq:trans_den}) becomes numerically unstable when the matrix becomes near-singular. In the latter situation, the destructive interference between the two diabats results in underestimated off-diagonal elements of the non-orthogonal diabatic Hamiltonian ($\vert H_{ab}^{\prime} \vert$), most likely due to high sensitivity of the XC functional to self-interaction error in this regime. When this situation causes a sign flip upon L\"owdin orthogonalization of the ALMO diabats using Eq.~(\ref{eq:msdft_2state}), i.e., $\vert H_{ab}^{\prime} \vert < \vert \mathcal{S}_{ab}(E_{a} + E_{b})/2 \vert$, the coupling value obtained becomes unreliable.

To address this challenge, we exploit an analogy to the generalized Slater-Condon rules to suggest an alternative that is applicable in the weak-coupling limit. As we noted previously, the MSDFT2 scheme can be viewed as a KS-DFT analogue to the generalized Slater-Condon rule \cite{Lowdin1950, Thom2009} for two non-orthogonal HF states (determinants) that have a non-zero overlap.\cite{Mao2019} Based on this realization, here we propose a KS-DFT analogue to the generalized Slater-Condon rule for two non-orthogonal determinants whose interstate overlap matrix possesses one vanishing singular value. 

We achieve this by first symmetrically orthogonalizing the occupied orbitals in two ALMO-based diabats, which yields $\bar{\mathbf{C}}_{\text{o}}^{(a)}$ and $\bar{\mathbf{C}}_{\text{o}}^{(b)}$. We then perform a singular value decomposition (SVD) on the interstate occupied orbital overlap matrix, $ \bar{\mathbf{S}}_{\text{oo}}^{(ab)}$,
\begin{equation}
\bar{\mathbf{S}}_{\text{oo}}^{(ab)} \equiv (\bar{\mathbf{C}}^{(a)}_{\text{o}})^{T} \mathbf{S} \bar{\mathbf{C}}^{(b)}_{\text{o}} = \mathbf{U}\mathbf{s}\mathbf{V}^{T},
\end{equation}
and then use the $\mathbf{U}$ and $\mathbf{V}$ matrices obtained to transform the orthogonal orbitals $\bar{\mathbf{C}}_{\text{o}}^{(a)}$ and $\bar{\mathbf{C}}_{\text{o}}^{(b)}$ to the L\"owdin-paired orbitals \cite{Lowdin1950}, $\tilde{\mathbf{C}}_{\text{o}}^{(a)} = \bar{\mathbf{C}}_{\text{o}}^{(a)} \mathbf{U}$ and $\tilde{\mathbf{C}}_{\text{o}}^{(b)} = \bar{\mathbf{C}}_{\text{o}}^{(b)} \mathbf{V}$. We proceed to order the L\"owdin-paired orbitals orbitals using their singular values such that the first vector in each set of orbitals ($\tilde{\mathbf{C}}_{\text{o}}^{(a)}$ and $\tilde{\mathbf{C}}_{\text{o}}^{(b)}$) corresponds to the pair of orbitals that overlap the least (associated with the smallest singular value, $s_1$), denoted here as $\tilde{\mathbf{C}}_{\text{o},1}^{(a)}$ and $\tilde{\mathbf{C}}_{\text{o},1}^{(b)}$. The remaining $N-1$ vectors in each set, $\tilde{\mathbf{C}}_{\text{o}, N-1}^{(a)}$ and $\tilde{\mathbf{C}}_{\text{o}, N-1}^{(b)}$, correspond to the remaining occupied orbitals (where $N$ denotes the total number of electrons) and $\mathbf{s}_{N-1}$ corresponds to the set of $N-1$ singular values. 
This results in two re-ordered sets of L\"owdin-paired orbitals: $\tilde{\mathbf{C}}_{\text{o}}^{(a)} = [\tilde{\mathbf{C}}_{\text{o},1}^{(a)}, \tilde{\mathbf{C}}_{\text{o}, N-1}^{(a)}]$ and $\tilde{\mathbf{C}}_{\text{o}}^{(b)} = [\tilde{\mathbf{C}}_{\text{o},1}^{(b)}, \tilde{\mathbf{C}}_{\text{o}, N-1}^{(b)}]$.

With this notation in place, we obtain the electronic coupling between diabats $\ket{\psi_{a}}$ and $\ket{\psi_{b}}$ as,
\begin{align}
H_{ab} &= \tilde{S}_{ab} \mathbf{P}_{1}^{(ab)} \cdot \mathbf{F}_{\text{KS}}^{(ab)} [\mathbf{P}_{N-1}^{(ab)}] \notag \\
&= \tilde{S}_{ab} \mathbf{P}_{1}^{(ab)} \cdot \left[\mathbf{h} + \mathbf{\RNum{2}} \cdot \mathbf{P}_{N-1}^{(ab)} + \mathbf{V}_{\text{xc}}[\tilde{\mathbf{P}}_{N-1}^{(ab)}] \right],
\label{eq:fock_approach}
\end{align}
where
\begin{align}
\mathbf{P}_{1}^{(ab)} &= (\tilde{\mathbf{C}}_{\text{o}, 1}^{(a)}) (\tilde{\mathbf{C}}_{\text{o}, 1}^{(b)})^T,  \\
\mathbf{P}_{N-1}^{(ab)} &= (\tilde{\mathbf{C}}_{\text{o}, N-1}^{(a)}) \mathbf{s}_{N-1}^{-1} (\tilde{\mathbf{C}}_{\text{o}, N-1}^{(b)})^T
\end{align}
are two transition density matrix-like objects, $\tilde{\mathbf{P}}_{N-1}^{(ab)} = (\mathbf{P}_{N-1}^{(ab)} + \mathbf{P}_{N-1}^{(ba)})/2 $ is the symmetrized version of $\mathbf{P}_{N-1}^{(ab)}$ and $\tilde{S}_{ab} = \prod_{i\neq 1}^{N} s_i$ is the reduced interstate overlap. The KS Fock matrix ($\mathbf{F}_{\text{KS}}$) is constructed from the transition density matrix between $N-1$ pairs of orbitals, which consists of contributions from the core-Hamiltonian ($\mathbf{h}$), two-electron integrals ($\mathbf{\RNum{2}}$), and the exchange-correlation potential ($\mathbf{V}_{\text{xc}}$) as shown in the second line of Eq.~(\ref{eq:fock_approach}). 

Our prescription in Eq.~(\ref{eq:fock_approach}) for calculating the diabatic coupling in the weak-coupling regime is a KS-DFT analogue to the generalized Slater-Condon rule for two HF determinants whose interstate occupied orbital overlap, $\mathbf{S}_{\text{oo}}^{(ba)}$, has one zero singular value\cite{Lowdin1950,Thom2009}
\begin{equation}
H_{ab} = \tilde{S}_{ab} \mathbf{P}_{1}^{(ab)} \cdot \left[\mathbf{h} + \mathbf{\RNum{2}} \cdot \mathbf{P}_{N-1}^{(ab)}  \right],
\end{equation}
where $\mathbf{\RNum{2}}$ denotes the contribution from the Coulomb interaction and full HF exchange. The approximation central to the application of our KS-DFT analogue to this Slater-Condon rule is that one can employ Eq.~(\ref{eq:fock_approach}) when the overlap between orbitals $\tilde{\mathbf{C}}_{\text{o},1}^{(a)}$ and $\tilde{\mathbf{C}}_{\text{o},1}^{(b)}$ is small but nonzero, which is the case in the weak-coupling regime.

In the following we refer to this approach to evaluate the diabatic coupling in the weak-coupling regime given in Eq.~(\ref{eq:fock_approach}) as the MSDFT2-wc scheme. We apply MSDFT2-wc in the cases discussed above where either orbital overlap is vanishingly small or destructive interference between the diabats occurs. Our identification of the weak-coupling regime in which Eq.~(\ref{eq:fock_approach}) is used to compute the diabatic coupling is based on the magnitude of the smallest singular value in $\mathbf{s}$ and a ``sign-flipping" criterion and is described in detail in Sec.~\ref{sec:comput_detail}.

\section{Implementation and Computational Details} \label{sec:comput_detail}

\subsection{Implementation of $\Delta$-ALMO(MSDFT2)}

We extended our previous implementation of ALMO(MSDFT2) \cite{Mao2019} in Q-Chem \cite{Shao2015} to construct the diabats involved in photoinduced ET processes using $\Delta$-ALMO(MSDFT2). To construct the LE diabat $\ket{D^\ast A}$, one first obtains the ground-state SCF solutions for both fragments. A non-aufbau electronic configuration of the donor is then prepared by swapping the occupied and virtual orbitals to reflect the nature of the excitation that is being constructed. As shown in Fig.~\ref{fig:almo_msdft_scheme}, one first computes the $\Delta$SCF solution for the excited donor ($D^\ast$) and then uses the orbitals on $D^\ast$ and ground-state $A$ to construct the initial guess for the full-system LE state (Eq.~(\ref{eq:almo_le_init})), which is followed by a variational optimization via the ALMO-$\Delta$SCF procedure introduced in Sec.~\ref{subsec:almo_diabats}. To prevent the collapse onto the ground electronic state, the IMOM method\cite{Barca2018} was employed in combination with the DIIS algorithm \cite{Pulay1982} in both the fragment $\Delta$SCF and the global system ALMO-$\Delta$SCF calculations. The CT states involved in this work correspond to the transition from the donor's HOMO to the acceptor's LUMO, and therefore were all constructed from ground-state unrestricted ALMO calculations. We note that the LE and CT states in this work are spin-symmetry broken, i.e., the commonly adopted spin-projection scheme\cite{Ziegler1977} for singlet states in $\Delta$SCF is not applied. Since spin-projected (LE and CT diabatic) states are no longer variationally optimized and the same broken-symmetry states were used in the previous studies using the MOM or IMOM methods and provided good accuracy \cite{Gilbert2008, Barca2018}, the choice to work with spin-symmetry broken states provides a convenient way to compute LE-CT couplings which would require the evaluation of additional matrix elements if spin-projection were employed.

The MSDFT and MSDFT2 coupling schemes, which we have implemented in the released version of the Q-Chem 5.3 package,\cite{Shao2015} are as introduced in our previous work\cite{Mao2019}. To evaluate the off-diagonal element of the non-orthogonal diabatic Hamiltonian ($H_{ab}^{\prime}$), our implementation first orthogonalizes the occupied orbitals constituting two ALMO diabats separately and then transforms these orbitals to be biorthogonal using L\"owdin's orbital pairing scheme,\cite{Lowdin1950} allowing one to identify the pair of orbitals having the smallest overlap. Our implementation identifies the weak-coupling regime based on two criteria: when the smallest singular value of the interstate overlap is smaller than $10^{-4}$ and/or when ``sign-flipping'' occurs in Eq.~(\ref{eq:msdft_2state}), i.e., when $\vert H_{ab}^{\prime} \vert < \vert \mathcal{S}_{ab}(E_{a} + E_{b})/2 \vert$. When either of these conditions occurs, we employ the MSDFT2-wc scheme given in Eq.~(\ref{eq:fock_approach}) to evaluate the diabatic coupling. We have employed the standard MSDFT2 scheme that uses the full sets of L\"owdin-paired orbitals from both diabats ($\tilde{\mathbf{C}}_{\text{o}}^{(a)}$ and $\tilde{\mathbf{C}}_{\text{o}}^{(b)}$) to construct the transition density matrix in all other cases. For calculations based on MSDFT, we have employed the original prescription given in Eq.~(\ref{eq:msdft1}) in all scenarios.

\subsection{Methods based on adiabatic-to-diabatic transformation}

To diabatize the adiabatic states obtained from TDDFT calculations, we employed the generalized Mulliken-Hush (GMH) \cite{Cave1996, Cave1997} and fragment charge difference (FCD) \cite{Voityuk2002} approaches. These methods first require one to select a set of adiabatic states that are relevant to a given electron transfer process. One then constructs the adiabatic-to-diabatic transformation by defining diabatic states as eigenstates of a given molecular property: dipole moment in GMH and difference in fragment charge populations in FCD. 

The GMH method utilizes the dipole operator\cite{Cave1996,Cave1997}, which requires calculation of both the dipole moment of each adiabat and the transition dipole between each pair of them. When only two adiabats (denoted with indices ``1'' and ``2'') are considered, the coupling between two resulting diabats (denoted with indices ``$a$'' and ''$b$'') is given by
\begin{equation}
H_{ab} = \frac{\vert \hat{\mu}_{12} \vert (E_2 - E_1)} {[(\boldsymbol{\mu}_{1} - \boldsymbol{\mu}_{2})^{2} + 4  \hat{\mu}_{12}^2]^{1/2}} ,
\label{eq:2state_gmh}
\end{equation}
where $E_{1}$, $E_{2}$ are the energies of the two adiabats, $\boldsymbol{\mu}_{1}$,  $\boldsymbol{\mu}_{2}$ are their dipole moments, and $\hat{\mu}_{12}$ denotes the projection of the transition dipole ($\boldsymbol{\mu}_{12}$) along the charge-transfer direction: $\hat{\mathbf{e}}_{\text{CT}} = (\boldsymbol{\mu}_{1} - \boldsymbol{\mu}_{2})/\vert \boldsymbol{\mu}_{1} - \boldsymbol{\mu}_{2} \vert$. 

The FCD method\cite{Voityuk2002}, on the other hand, defines the diabats as eigenstates of the fragment charge difference matrix ($\Delta Q_{ij} = Q_{ij}(D) - Q_{ij}(A)$), where $D$ denotes the donor and $A$ the acceptor). The calculation of the diagonal and off-diagonal elements of $\boldsymbol{\Delta}\mathbf{Q}$ requires performing charge population analysis with the electron density of each adiabat and the transition density between each pair of them. In the 2-state case, the diabatic coupling is given by
\begin{equation}
H_{ab} = \frac{ \vert \Delta Q_{12} \vert (E_2 - E_1)}{\sqrt{(\Delta Q_{11} - \Delta Q_{22})^2 + 4 \Delta Q_{12}^2}}.
\label{eq:2state_fcd}
\end{equation}
Here we evaluated the differences in fragment charge populations subject to the Mulliken charge population scheme.\cite{Mulliken1955}

The reference values for the diabatic couplings were obtained from GMH diabatization of EOM-CCSD states. Unless otherwise specified, we used the 2-state version of the GMH and FCD diabatization schemes given by Eqs.~(\ref{eq:2state_gmh}) and (\ref{eq:2state_fcd}). The procedures for the 3-state GMH diabatization required for the indole-guanine complex and FCD for a general number of states that was used to diabatize the TDDFT adiabats for the pentacene dimer are described in SI Secs.~S1 and S2, respectively.

\subsection{Computational setup}

The calculations in this work were all performed with our locally developed version of the Q-Chem 5.3 package.\cite{Shao2015} Unless otherwise specified, all the DFT-based calculations for diabatic couplings (based on ALMO or TDDFT) were performed with the $\omega$B97X-D functional \cite{Chai2008} and the 6-31+G(d) basis set \cite{Hehre1972, Frisch1984} on a (99, 590) grid, which has 99 radical shells and 590 Lebedev points in each. To test the performance of each method on the choice of functional, we also generated results for the indole-guanine complex (Sec.~\ref{subsec:Ind_G}) with pure GGA (BLYP,\cite{Becke1988, Lee1988} PBE \cite{Perdew1996}), global hybrid (B3LYP,\cite{Becke1993} PBE0 \cite{Adamo1999}), and two other range-separated hybrid (RSH) functionals (CAM-B3LYP,\cite{Yanai2004} LRC-$\omega$PBEh\cite{Rohrdanz2009}). Using the same system, we also investigated the basis set dependence of the diabatic couplings calculated from ALMO- and TDDFT-based methods and revealed that the results are largely insensitive to the choice of basis set (see Fig.~S1 in the SI). Both the standard and ALMO-based SCF calculations were converged to a DIIS error below 10$^{-8}$ a.u. The linear-response TDDFT calculations were performed within the Tamm-Dancoff approximation \cite{Hirata1999} and only the singlet excited states were considered. The reference values for the diabatic couplings were obtained from diabatization of the adiabatic states obtained from EOM-CCSD \cite{Stanton1993,Krylov2008} calculations, where the IP (ionization potential) variant was applied to the indole-guanine complex while the EE (electronic excitation) variant was applied to the others. Both EOM-CCSD variants employ a closed-shell reference for the systems that we investigate here and Hartree-Fock (HF) theory was used as the default method to calculate the reference state. While it is often possible to identify the characters of excited states using the canonical orbitals from HF calculations, allowing one to identify the excited states relevant to a photoinduced ET process, the HF virtual orbitals of the 1-naphthol--CHCl$_3$ complex (Sec.~\ref{subsec:naphthol_chcl3}) are highly diffuse, rendering it difficult to identify the relevant states contributing to the ET process. To address this issue, we generated the reference orbitals from a ground-state SCF calculation using the LRC-$\omega$PBEh functional for this system, which resulted in canonical KS orbitals that allow for transparent assignment of relevant excited states.

The geometries of the systems used here were obtained from previous studies: the cationic indole--guanine complex from Ref.~\citenum{Voityuk2013}, naphthalene--tetracyanoethylene complex from Ref.~\citenum{Stein2009}, 1-naphthol--CHCl$_3$ complex from Ref.~\citenum{Chaudhuri2019}, and the pentacene dimer from Ref.~\citenum{Zeng2014}. The scan of inter-monomer distances and monomer rotation angles for the pentacene dimer was performed with respect to the ``reference'' dimer structure given in Ref.~\citenum{Zeng2014}, in which $d_{\text{CC}}$ (shown in the inset of Fig.~\ref{fig:pentacene_dimer_scan}) is 5.98 \AA. For completeness we provide these geometries in a ZIP file available in the SI.

\section{Results} \label{sec:results}

Here we benchmark a set of excited-state electron and hole transfer systems to demonstrate the performance of $\Delta$-ALMO(MSDFT2) in obtaining the diabatic couplings involved in these photoinduced processes by comparing to the reference values obtained from a GMH diabatization of states computed at the EOM-CCSD level. Because TDDFT is commonly employed to calculate adiabatic excited states, we also compare our $\Delta$-ALMO(MSDFT2) results to diabatic couplings obtained from the application of adiabatic-to-diabatic (ATD) transformation schemes (GMH and FCD) to the TDDFT states. The four systems on which we benchmark these methods were chosen to highlight four fundamentally important applications of excited-state electron and hole transfer as well as the specific challenges associated with their diabatization. 

In particular, we first consider ground- and excited-state hole transfer between indole and guanine, which is relevant to DNA repair mechanisms and where the approaches based on the ATD transformation require more than two states to obtain accurate results. We then treat photoexcited electron transfer in naphthalene--tetracyanoethylene, which mimics the photoinduced charge separation process within a donor-acceptor dyad in organic electronics and illustrates a case where $\Delta$-ALMO(MSDFT2) can be used when the LE state involves more than one orbital transitions. Next we focus on photoexcited electron transfer in the 1-naphthol--CHCl$_3$ complex, which serves as a paradigm of chromophore-to-solvent electron transfer and demonstrates the need to use the MSDFT2-wc coupling scheme introduced in Sec.~\ref{subsec:fock_approach} in the weak-coupling regime. Finally we address the diabatic coupling between the LE and CT states in the pentacene dimer, which is relevant for the superexchange mechanism that has been shown to be essential for efficient singlet fission \cite{Chan2013,Berkelbach2013b,Zeng2014} and presents a challenge for the commonly used schemes based on ATD transformation such as GMH due to the involvement of multiple adiabatic excited states and lack of a uniform charge-transfer direction.

\subsection{Cationic Indole-Guanine complex} \label{subsec:Ind_G}

\begin{figure}[h!]
	\centering
	\includegraphics[width=0.48\textwidth]{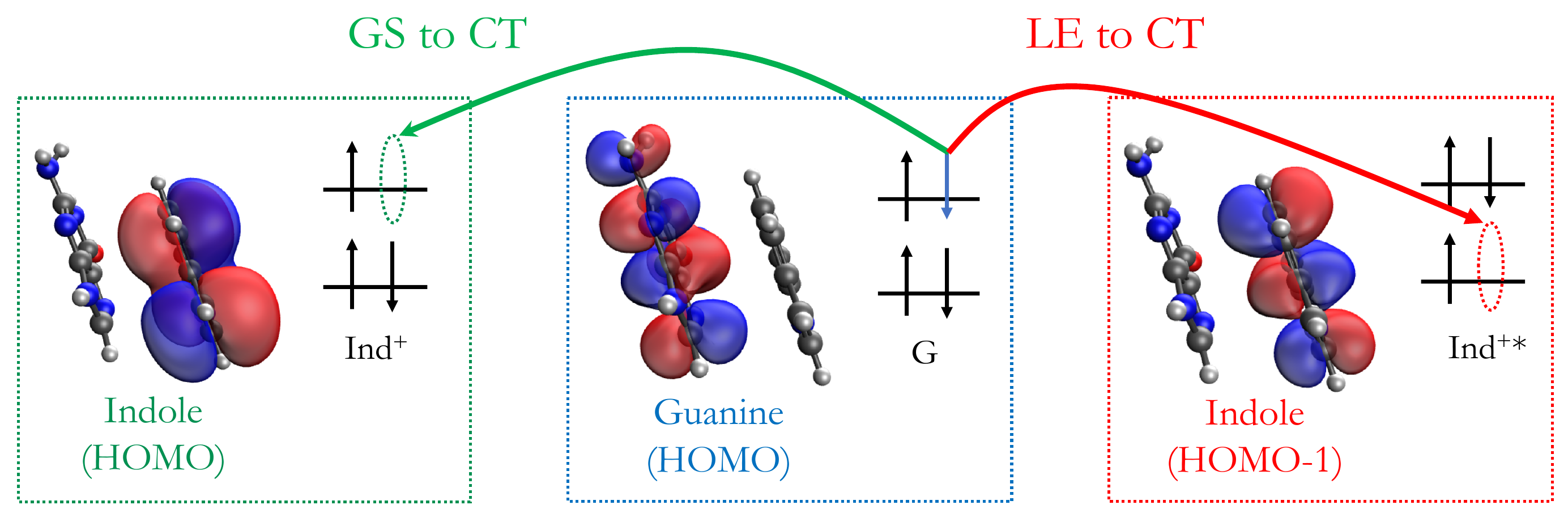}
	\caption{The ground- and excited-state hole transfer processes in the cationic indole-guanine ([Ind-G]$^{+}$) complex. The hole is transferred from the HOMO (ground-state HT) or HOMO-1 (excited-state HT) of indole to the HOMO of guanine. These orbitals are obtained from ALMO-based diabats for Ind$^+$-G and Ind$^{+\ast}$-G and visualized with an isovalue of 0.02~a.u. }
	\label{fig:ind_guanine_HT}
\end{figure}

The cationic indole-guanine complex, [Ind-G]$^{+}$, serves as a prototypical example of hole transfer from nucleobases to amino acid residues in DNA-protein complexes, a process that can inhibit the oxidative damage of DNA.\cite{Butchosa2012, Voityuk2013} Here we consider two hole transfer processes from indole, whose HOMO and HOMO$-$1 orbitals are close in energy, to guanine. These two processes are illustrated in Fig.~\ref{fig:ind_guanine_HT} where for the ground state process the hole is on the HOMO of indole (Ind$^+$) and for the excited state process it is on the HOMO$-$1 (Ind$^{+\ast}$). We note that physically the reverse process, i.e., hole transfer from guanine to indole, is relevant to DNA protection. However, considering the forward process allows us to use the language that aligns with that of Sec.~\ref{subsec:almo_diabats}, with the Ind$^+$-G and Ind$^{+\ast}$-G configurations assigned as the ground (GS) and locally excited (LE) initial states, respectively, and their diabatic couplings to the CT state denoted as $H_{\text{GS-CT}}$ and $H_{\text{LE-CT}}$. Since the diabatic states involved in the forward and reverse processes are identical, the diabatic couplings are the same.

\begin{figure}[t!]
	\centering
	\includegraphics[width=0.48\textwidth]{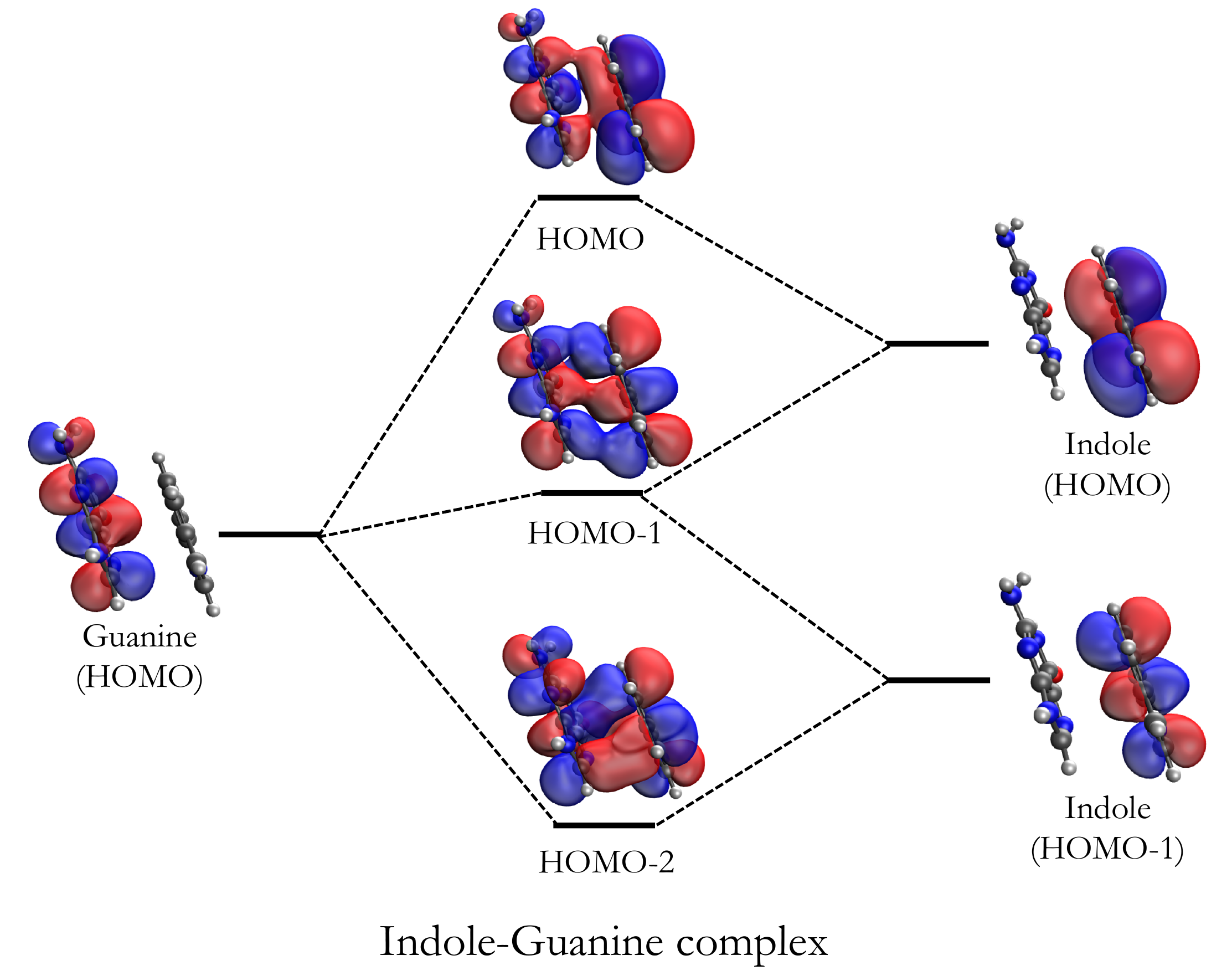}
	\caption{Depiction of the connection between the adiabatic and diabatic molecular orbitals involved in the ground- and excited-state hole transfer in the cationic [Ind-G]$^{+}$ complex. The adiabatic orbitals that are plotted (shown in the middle) are obtained from the closed-shell HF reference for the EOM-IP-CCSD calculation, and the diabatic orbitals plotted (shown on the two sides and also in Fig.~\ref{fig:ind_guanine_HT}) from ALMO-based diabats for Ind$^+$-G and Ind$^{+\ast}$-G. The orbitals are visualized with an isovalue of 0.02 a.u.}
	\label{fig:ind_guanine_orbs}
\end{figure}

To obtain the reference values for the GS-CT and LE-CT couplings, we first calculated the three lowest-lying adiabatic states for the cationic system using EOM-IP-CCSD and then performed GMH diabatization on them (see Sec.~\ref{sec:comput_detail}). The EOM-IP-CCSD calculation starts from the closed-shell, charge-neutral reference system. The three lowest-lying resultant adiabatic states are dominated by ionization from the highest three occupied orbitals (HOMO, HOMO$-$1, HOMO$-$2) of the global system. These orbitals, as shown in the middle column of Fig.~\ref{fig:ind_guanine_orbs}, are approximately linear combinations of orbitals in the diabatic picture (ALMOs), namely, the HOMO and HOMO-1 of indole and the HOMO of guanine. The delocalized nature of the adiabatic orbitals and in particular the entangled character of HOMO$-$1 of the global system (in which all three diabatic orbitals possess non-negligible weights) make it necessary to invoke 3-state GMH diabatization \cite{Rust2002} rather than using the more commonly used 2-state approximation (see Secs.~S1 and S3.A in the SI for the detailed procedure). The reference values obtained are $H_{\text{GS-CT}} = 277$~meV and $H_{\text{LE-CT}} = 60$~meV.

\begin{figure*}[t!]
	\centering
	\includegraphics[width=0.75\textwidth]{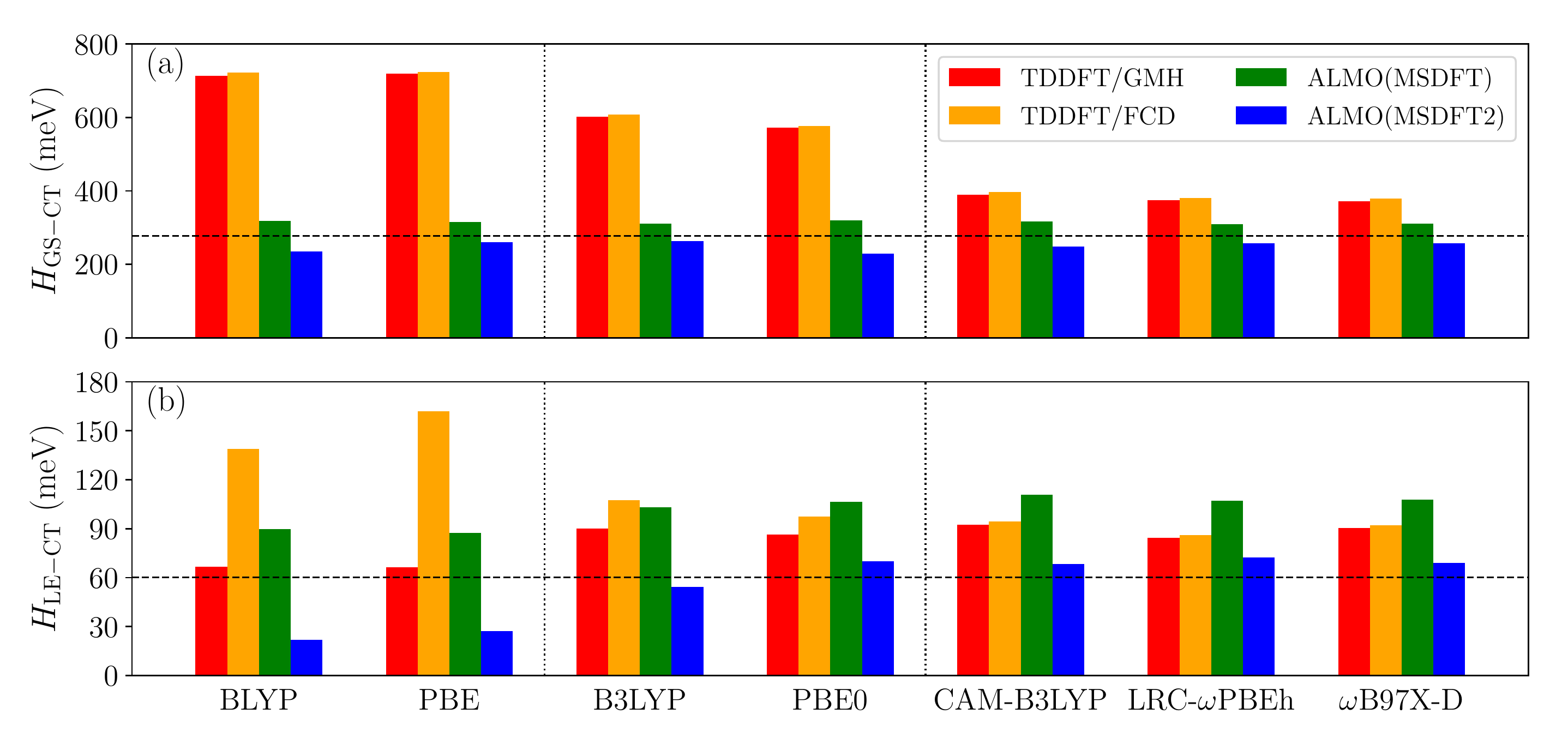}
	\caption{Diabatic couplings for the (a) ground and (b) excited state hole transfer in the [Ind-G]$^{+}$ complex calculated using TDDFT with the 3-state GMH and FCD diabatization schemes as well as with two ALMO-based approaches. The calculations were performed with density functionals ranging from pure GGA to global and range-separated hybrid functionals. The horizontal dashed lines mark the reference values for $H_{\text{GS-CT}}$ and $H_{\text{LE-CT}}$, obtained at the EOM-IP-CCSD/6-31+G(d) level, and the vertical dotted lines indicate the grouping of pure GGA, global hybrid, and range-separated hybrid functionals.}
	\label{fig:ind_guanine_gsex}
\end{figure*}

Since TDDFT is commonly used to investigate excited-state processes we also calculated the GS-CT and LE-CT couplings for the HT in this system using TDDFT followed by 3-state GMH and FCD diabatization. Unlike EOM-IP-CCSD, the TDDFT calculations employ the unrestricted KS-DFT description of the cationic complex as the reference. With global hybrid (B3LYP and PBE0) and range-separated hybrid (RSH) functionals (CAM-B3LYP, LRC-$\omega$PBEh, and $\omega$B97X-D), the first TDDFT excited state is dominated by a local excitation involving a transition from indole's HOMO$-$1 to HOMO, while the second excited state exhibits marked charge transfer from guanine to indole. We assign the LE and CT characters to these states using the detachment and attachment densities \cite{HeadGordon1995} associated with them, as shown in SI Fig.~S2. When pure GGA functionals (BLYP and PBE) are employed, the state featuring G$\rightarrow$Ind CT is shifted to become the 4th excited state.

The resulting GS-CT and LE-CT couplings obtained from applying the adiabatic-to-diabatic transformation schemes, GMH and FCD, to the TDDFT states are shown in Fig.~\ref{fig:ind_guanine_gsex}. For the GS-CT coupling shown in Fig.~\ref{fig:ind_guanine_gsex}(a), TDDFT/GMH and TDDFT/FCD both yield similar results that systematically overestimate $H_{\text{GS-CT}}$. The relative errors are over 100\% when paired with pure and global hybrid functionals and reduce to 30--40\% when RSH functionals are employed. While Fig.~\ref{fig:ind_guanine_gsex}(b) demonstrates that GMH and FCD also give similar results for the LE-CT coupling when using RSH and global hybrid functionals with relative errors of 40--50\%, they differ dramatically when paired with the two GGA functionals. In particular, when using BLYP or PBE, TDDFT/GMH gives more accurate LE-CT couplings, likely due to fortuitous error cancellation, whereas TDDFT/FCD significantly overestimates the coupling by $\sim$150\%. Hence, diabatization based on TDDFT excited states is ill-suited for treating hole transfer in this explicitly charged system.

In contrast to the large overestimation of TDDFT combined with GMH and FCD schemes, ALMO(MSDFT2) gives accurate results (with the lowest error being 5\% and the largest 17\%) for the diabatic coupling of the GS-CT process, even when combined with pure GGA functionals. This observation is consistent with our previous work\cite{Mao2019} where we observed for a range of ground-state ET and HT processes that ALMO(MSDFT2) could give highly accurate results even when combined with lower-tier functionals. Whereas the $\Delta$-ALMO(MSDFT2) scheme that we have introduced here suffers from poor performance for the LE-CT coupling when paired with pure functionals producing an error of 50--60\%, it yields accurate results when combined with global hybrid and RSH functionals, with relative errors of 10--20\% that are significantly lower than those obtained using the other methods, except for TDDFT/GMH when combined with GGA functionals, which has very likely benefited from fortuitous cancellation of error. In contrast to MSDFT2, using the MSDFT coupling scheme leads to systematic overestimation of both the GS-CT and LE-CT diabatic couplings when combined with any tier of functional, whose results degrade markedly in the excited-state process with errors of 45--85\%. The systematic overestimation of MSDFT couplings is in agreement with the trend revealed by our previous ground-state ET and HT benchmarks.\cite{Mao2019} Thus, when combined with global hybrid or RSH functionals ALMO(MSDFT2) provides a good balance of general applicability and accuracy in obtaining the diabatic couplings for the ground- and excited-state hole transfer processes in this system.

\subsection{Naphthalene-tetracyanoethylene complex}

\begin{figure*}[t!]
	\centering
	\includegraphics[width=0.7\textwidth]{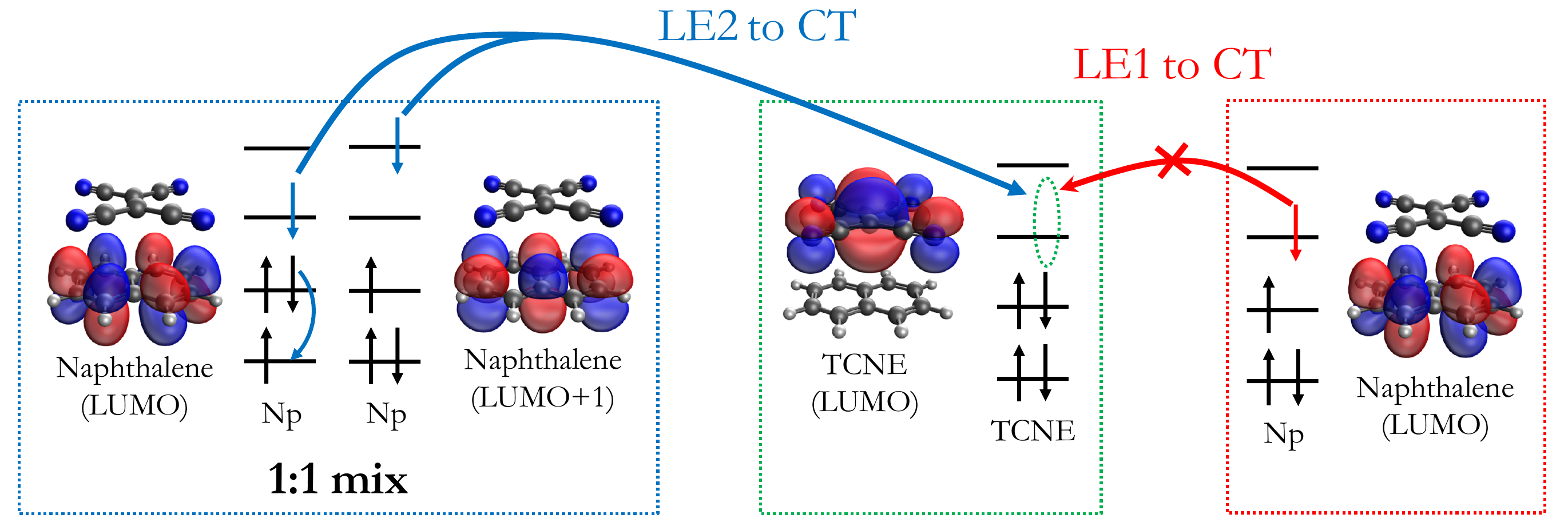}
	\caption{The excited-state electron transfer processes in the naphthalene-tetracyanoethylene (Np-TCNE) complex. The electron can be transferred from Np in the LE1 or  LE2 locally excited state to TCNE giving rise to a CT state corresponding to an electron transfer from Np's HOMO to TCNE's LUMO. The former process is symmetry-forbidden while the latter is symmetry-allowed. The LE1 state is dominated by the HOMO$\rightarrow$LUMO transition. The LE2 state consists of a 1:1 mix of the HOMO$-$1$\rightarrow$LUMO and HOMO$\rightarrow$LUMO$+$1 transitions, which has non-zero coupling with the CT state.}
	\label{fig:naph_tcne_orbital}
\end{figure*}

The naphthalene-tetracyanoethylene (Np-TCNE) complex is a prototypical model system for a donor-acceptor dyad resembling those in photovoltaic devices. In contrast with the [Ind-G]$^{+}$ complex investigated above, this complex possesses a charge-neutral, closed-shell ground state. The naphthalene molecule has two low-lying singlet excited states which we denote as LE1 and LE2 (also denoted as $^{1}\mathrm{L}_a$ and $^{1}\mathrm{L}_b$ states in literature,\cite{Platt1949} respectively).
A TDDFT calculation for naphthalene using an RSH functional $\omega$B97X-D reveals that the LE1 state is dominated by the HOMO$\rightarrow$LUMO transition while the LE2 state a 1:1 mix of the HOMO$-$1$\rightarrow$LUMO and HOMO$\rightarrow$LUMO$+$1 transitions as illustrated in Fig.~\ref{fig:naph_tcne_orbital}. The calculations reveal that the two states are 4.66 and 4.92 eV above the ground state, respectively. The charge transfer excitation from the HOMO of naphthalene to the LUMO of TCNE has a lower energy (2.6 eV) \cite{Stein2009} and thus can be accessed via photoinduced electron transfer from the LE states on naphthalene. 

The equilibrium geometry of this complex is of $C_{2v}$ symmetry and the inter-fragment distance (defined as the distance between the midpoints of the central \ce{C-C} bonds in each monomer) is 3.9 \AA. Owing to this symmetry, the LE1 state and the lowest CT state belong to the B$_2$ and B$_1$ irreducible representations, respectively, causing their coupling to vanish. This zero-coupling result is exactly reproduced (to numerical precision) by both the $\Delta$-ALMO(MSDFT) and $\Delta$-ALMO(MSDFT2) approaches. The LE2 state, on the other hand, belongs to the B$_1$ irreducible representation leading to a non-zero coupling with the CT state. Using EOM-EE-CCSD to compute the first two B$_2$ states followed by a 2-state GMH diabatization gives a reference value of 128 meV for the diabatic coupling between the LE2 and CT states.

\begin{figure}[h!]
	\centering
	\includegraphics[width=0.45\textwidth]{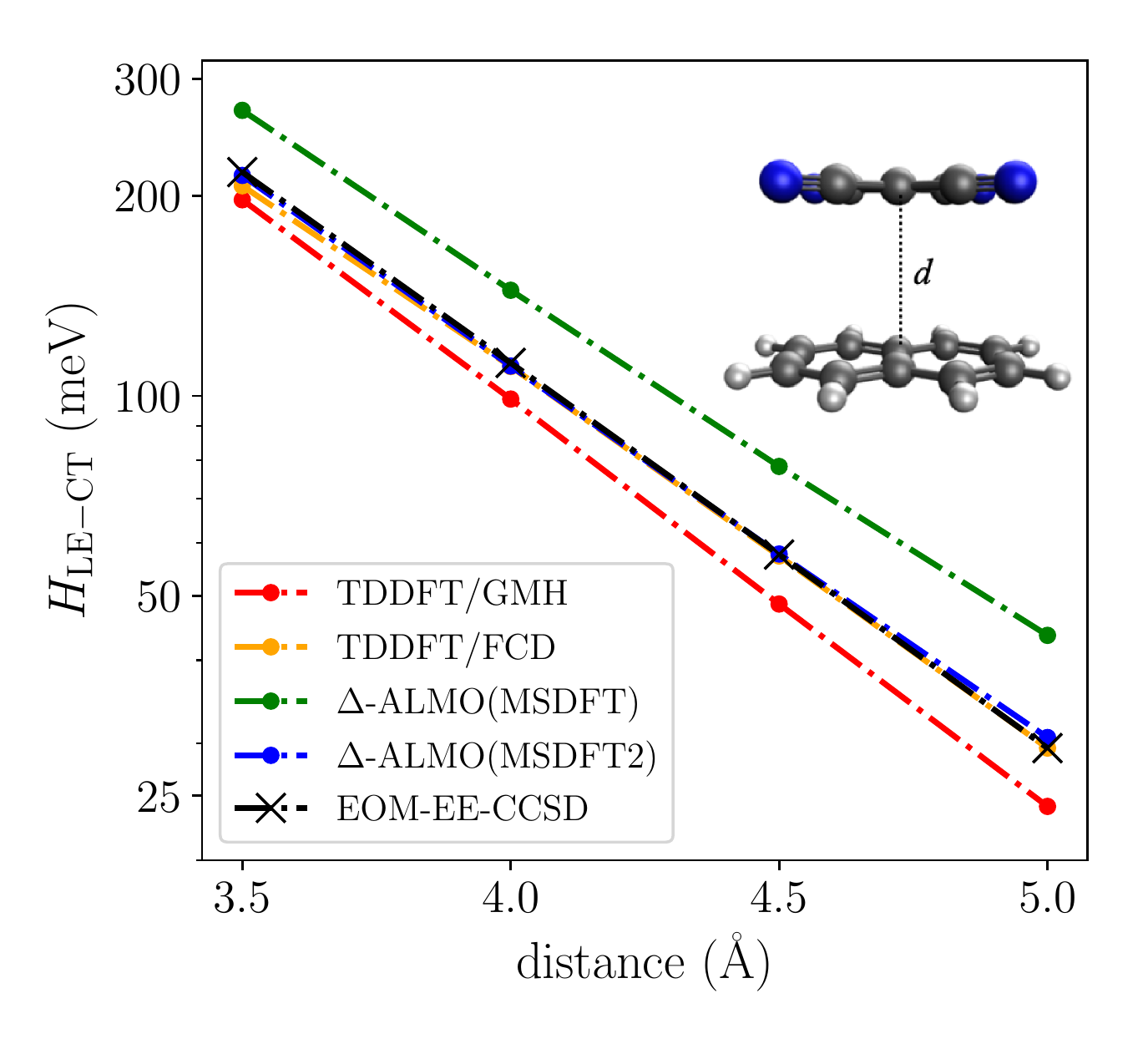}
	\caption{Distance dependence of the LE-CT coupling (in meV) for the photoinduced ET in the naphthalene-TCNE complex obtained from TDDFT and ALMO-based calculations. The LE state corresponds to the LE2 state of naphthalene and the CT state corresponds to the transition from naphthalene's HOMO to TCNE's LUMO. The calculations were performed with inter-fragment distances equal to 3.5, 4.0, 4.5, and 5.0 \AA\ and the reference values were obtained at the EOM-EE-CCSD/6-31+G(d) level.}
	\label{fig:naph_tcne_dist_scan}
\end{figure}

The LE2 state of Np, in principle, cannot be described by standard $\Delta$SCF-based methods since it corresponds to a linear superposition of two equally weighted transitions: HOMO$\rightarrow$LUMO$+$1 and HOMO$-$1$\rightarrow$LUMO. However, exploiting the equal weights of these two configurations in the LE2 state, one can calculate their respective couplings with the CT state first, and then evaluate the coupling between the LE2 and CT states using
\begin{equation*}
H_{\text{LE-CT}} = [H(\mathrm{H-1\rightarrow L}, \text{CT}) + H(\mathrm{H\rightarrow L+1}, \text{CT})]/\sqrt{2}.
\end{equation*}
The resulting LE-CT couplings obtained by $\Delta$-ALMO(MSDFT) and $\Delta$-ALMO(MSDFT2) are 164 and 127 meV, respectively (see Table~S3 in the SI for the two intermediate couplings between a single ALMO-$\Delta$SCF state and the CT state) with the latter in excellent agreement with the EOM-EE-CCSD reference (128 meV).

We also obtained the results of TDDFT-based diabatization for the Np-TCNE complex at the equilibrium geometry (see SI Table~S4 for details about the obtained TDDFT states). Using the lowest two B$_1$ excited states as the adiabatic basis, the result of FCD diabatization (126 meV) also agrees very well with the EOM-EE-CCSD reference although TDDFT underestimates the energy of the lowest B$_1$ (CT) state by $\sim$1 eV (see Table~S4), while GMH underestimates the coupling slightly (114 meV). The improved accuracy of approaches based on the ATD transformation of TDDFT states for this closed-shell system are in stark contrast with the more inaccurate results obtained for the open-shell [Ind-G]$^{+}$ complex investigated in Sec.~\ref{subsec:Ind_G}.

We now investigate the ability of TDDFT- and ALMO-based approaches to capture the distance dependence of the LE-CT coupling over inter-fragment separations from 3.5 to 5.0 \AA\ with fixed monomer geometries. The results in Fig.~\ref{fig:naph_tcne_dist_scan} show that all methods examined are able to qualitatively capture the exponential decay of the LE-CT coupling, with $\Delta$-ALMO(MSDFT2) and TDDFT/FCD capturing the benchmark result within graphical accuracy. In contrast, $\Delta$-ALMO(MSDFT) systematically overestimates the coupling with the relative error growing from 24\% to 48\% as the inter-fragment distance increases, whereas TDDFT/GMH underestimates the coupling over the entire range with an (unsigned) relative error increasing from 9\% to 18\% with increasing distance. The increasing size of the relative error with increasing donor-acceptor separation implies that, in addition to inaccurately capturing the diabatic couplings, these methods also incorrectly capture the rate of decay of the couplings with donor-acceptor distance. In contrast, $\Delta$-ALMO(MSDFT2) accurately captures the diabatic couplings and their distance dependence for excited state ET in the Np-TCNE complex which forms a prototypical example of donor-acceptor dyads.

\subsection{Naphthol-CHCl$_3$ complex} \label{subsec:naphthol_chcl3}

The 1-naphthol-CHCl$_3$ complex is a prototypical system exhibiting chromophore-to-solvent electron transfer, a process that can play an important role in the non-radiative decay pathway of excited states in the condensed phase. The ultrafast photoinduced ET from 1-naphthol (NpOH) to CHCl$_3$ was recently investigated using Marcus theory combined with TDDFT/FCD diabatization\cite{Chaudhuri2019}. Here we examine the ability of $\Delta$-ALMO(MSDFT2) to capture the LE-CT couplings involved in this process. Importantly, this example also allows us to demonstrate how the $\Delta$-ALMO(MSDFT2) approach can be used in the weak coupling regime using the approaches introduced in Sec.~\ref{subsec:fock_approach}. 

A TDDFT calculation of the NpOH-CHCl$_3$ complex shows that the two lowest-lying singlet excited states are dominated by local excitations on NpOH: the first is dominated by the HOMO$\rightarrow$LUMO transition and the second by the HOMO$\rightarrow$LUMO+1 transition. In the following discussion, we denote these locally excited states LE1 and LE2, respectively. Unlike the Np-TCNE complex investigated above, the energy of the lowest CT state in this system, which corresponds to the transition from the HOMO of NpOH to the LUMO of CHCl$_3$, is of a higher energy than both of the LE states. Because the energy of the CT state predicted by EOM-EE-CCSD is markedly higher than those from TDDFT, which lifts the CT state up to the 7th singlet excited state of this system (with TDDFT it is the 3rd singlet excited state), it becomes necessary to include a relatively large number ($\sim$10) of states in the EOM-EE-CCSD calculation. The energies of these relevant adiabatic excited states given by TDDFT and EOM-EE-CCSD are provided in Table~S5 in the SI. 

\begin{figure}[h!]
	\centering
	\includegraphics[width=0.45\textwidth]{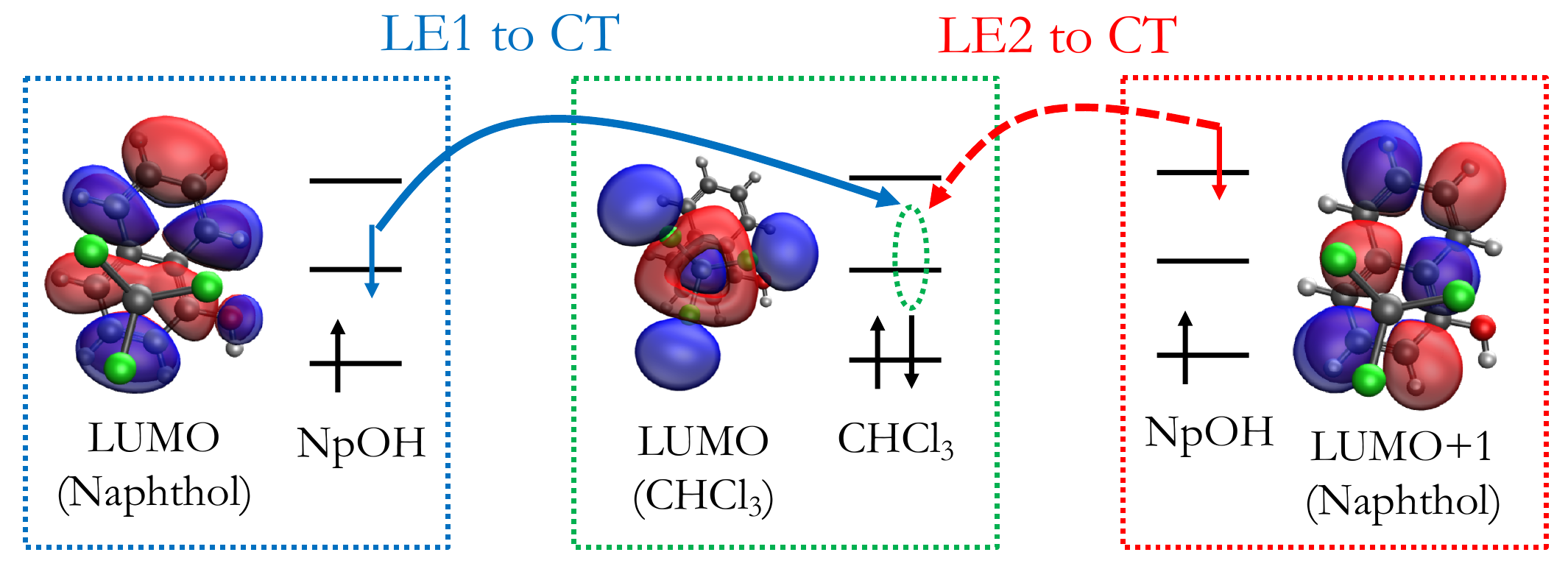}
	\caption{The excited state electron transfer processes between 1-naphthol (NpOH) and CHCl$_3$. The electron can be transferred from NpOH in two different locally excited states (LE1 and LE2) to CHCl$_3$ giving rise to a CT state that corresponds to the transition from the HOMO of NpOH to the LUMO of CHCl$_3$. The two LE states are dominated by the HOMO$\rightarrow$LUMO and HOMO$\rightarrow$LUMO+1 excitations, respectively. The LE2 state is weakly coupled to the CT state (indicated by the dashed arrow) due to the mismatched symmetry of NpOH's LUMO+1 and CHCl$_3$'s LUMO.}
	\label{fig:naphthol_chcl3_orbital}
\end{figure}

The two photoinduced ET processes considered in this work, corresponding to the LE1$\rightarrow$CT and LE2$\rightarrow$CT pathways, are illustrated in Fig.~\ref{fig:naphthol_chcl3_orbital}. The LE2$\rightarrow$CT pathway involves the LE2 state which is low-lying in energy (based on EOM-EE-CCSD calculations) but is only weakly coupled to the CT state due to the destructive overlap between NpOH's LUMO+1 and CHCl$_3$'s LUMO. In contrast, the LE1$\rightarrow$CT pathway involves the higher-energy LE1 state which is more strongly coupled with the CT state since the symmetry of the LUMOs of NpOH and CHCl$_3$ allows constructive orbital overlap (shown in Fig.~\ref{fig:naphthol_chcl3_orbital}). The reference values that we obtained using EOM-EE-CCSD/GMH for the LE1-CT and LE2-CT couplings are 72 and 17 meV, respectively.

\begin{figure}[h!]
	\centering
	\includegraphics[width=0.4\textwidth]{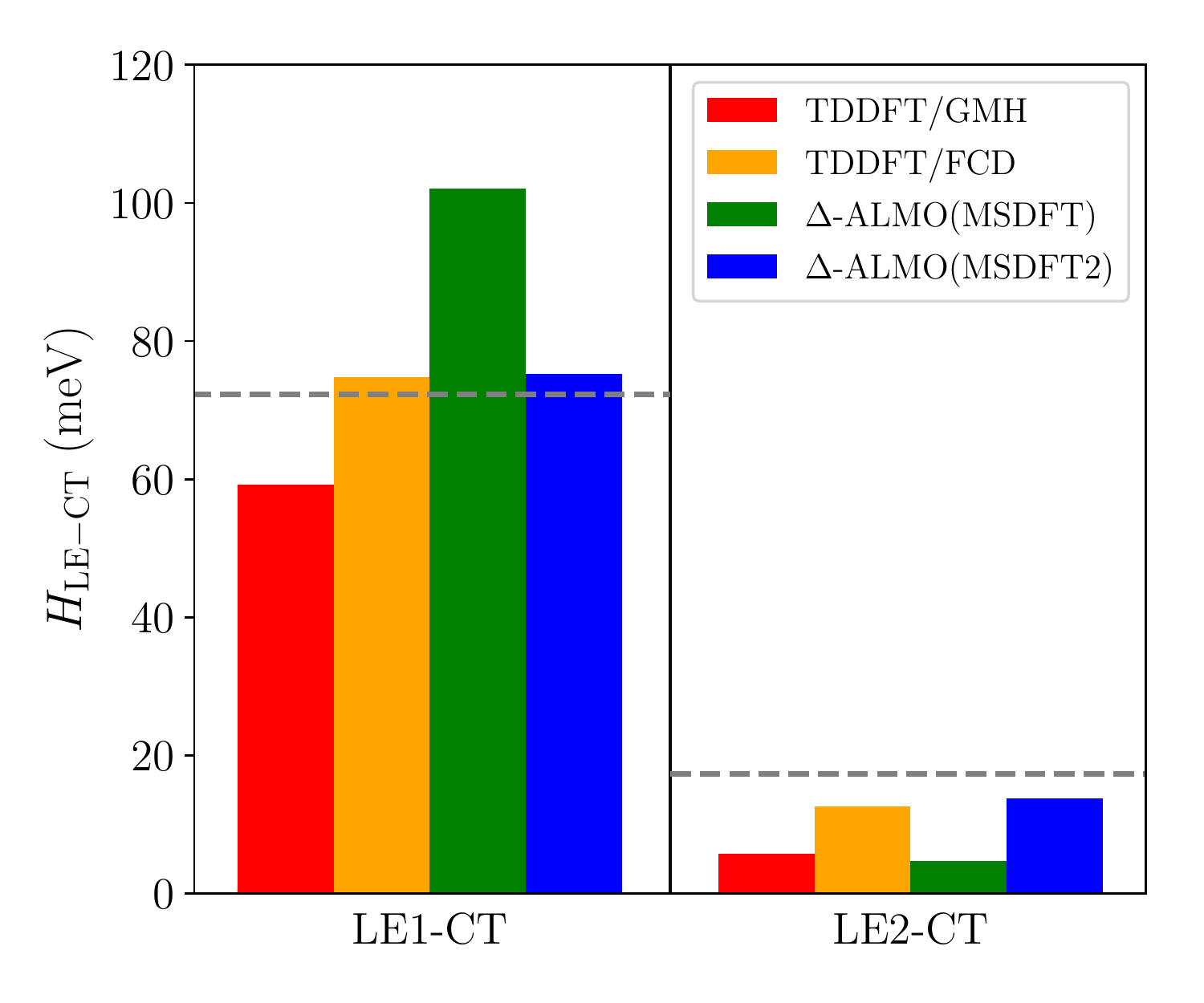}
	\caption{Diabatic couplings (in meV) between the two LE states on 1-naphthol and the CT state that corresponds to the transition from 1-naphthol's HOMO to CHCl$_3$'s LUMO evaluated with TDDFT and ALMO-based diabatization schemes. The reference values obtained from EOM-EE-CCSD/GMH calculations for these two couplings are shown in grey dashed lines.}
	\label{fig:naphthol_chcl3}
\end{figure}

We now turn to the performance of TDDFT- and ALMO-based schemes in predicting the LE1-CT and LE2-CT couplings, presented in Fig.~\ref{fig:naphthol_chcl3}.
The TDDFT/FCD results are in good agreement with the reference values in general while TDDFT/GMH underestimates both couplings with a relative error of 18\% for LE1-CT and 66\% for LE2-CT, mirroring the good performance of TDDFT/FCD and underestimation of TDDFT/GMH that we observed for Np-TCNE. $\Delta$-ALMO(MSDFT) overestimates the coupling between the LE1 and CT states with an error of 41\%, while our scheme, $\Delta$-ALMO(MSDFT2), again gives excellent agreement with the benchmark result, with only a relative error of 4\%, even though the broken-symmetry $\Delta$SCF treatment markedly underestimates the energy of the LE1 state (see Table~S5 in the SI).

As discussed in Sec.~\ref{subsec:fock_approach}, in the weak-coupling regime one has to adopt an alternative form of the coupling that we denote as MSDFT2-wc. In the case of the LE2-CT coupling in the NpOH-CHCl$_3$ complex, the weak coupling arises from the destructive overlap between NpOH's LUMO+1 and CHCl$_3$'s LUMO and is signaled by the smaller $\vert H_{ab}^{\prime} \vert$ than $\vert \mathcal{S}_{ab}(E_{a} + E_{b})/2 \vert$. Using the MSDFT-wc coupling prescription given in Eq.~(\ref{eq:fock_approach}) to evaluate the LE2-CT coupling leads to a value of 14 meV, whose error is within $\sim$20\% of the benchmark value (17 meV). Since $\Delta$-ALMO(MSDFT) is less prone to issues in the weak-coupling regime, the LE2-CT diabatic coupling can be generated using the original MSDFT coupling scheme in Eq.~(\ref{eq:msdft1}) but the resulting LE2-CT coupling is much smaller than the reference value, with a relative error of 73\%. These results suggest that the MSDFT2-wc approach that we proposed as a KS-DFT analogue to the generalized Slater-Condon rule for HF determinants whose interstate orbital overlap has one vanishingly small singular value in Sec.~\ref{subsec:fock_approach} provides a practical and fairly accurate approach in the weak-coupling regime arising from destructive orbital overlap in the NpOH-CHCl$_3$ complex.

\subsection{Pentacene dimer} \label{subsec:pentacene}

\begin{figure}[h!]
	\centering
	\includegraphics[width=0.4\textwidth]{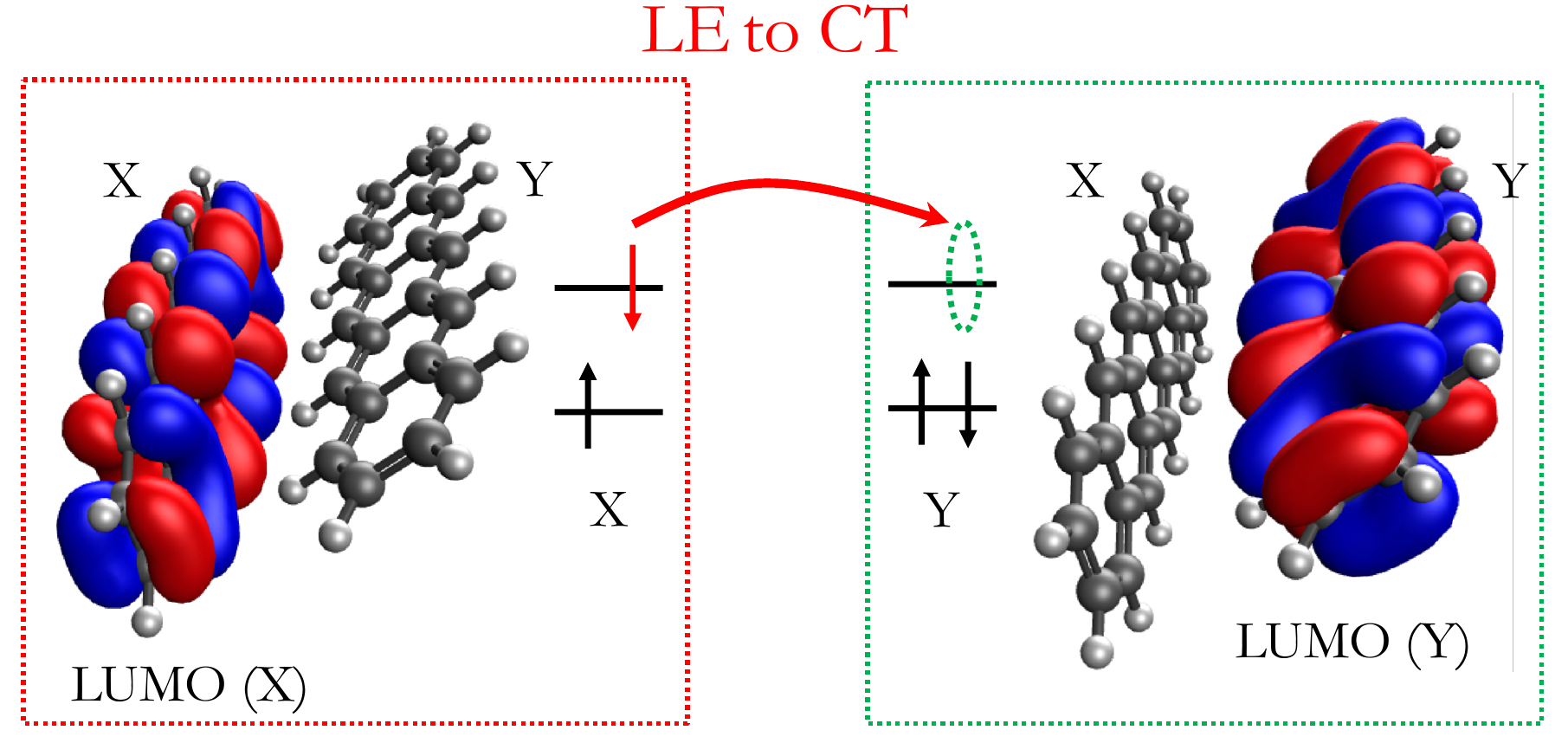}
	\caption{The change in the electronic configuration for the diabatic couplings investigated for the pentacene dimer. This change is equivalent to an electron transfer from monomer X in an LE state to monomer Y, giving rise to a charge-separated pentacene dimer (the CT state). The LE state corresponds to the HOMO$\rightarrow$LUMO transition on monomer X and the CT state corresponds to the transition from X's HOMO to Y's LUMO.}
	\label{fig:pentacene_dimer_orbs}
\end{figure}

Understanding the photophysics associated with the pentacene dimer plays a central role in mechanistic studies of singlet fission,\cite{Zimmerman2011,Berkelbach2013b,Feng2013,Zeng2014} where a singlet exciton is split into two triplet excitons and which provides a promising route for efficient solar energy conversion.\cite{Smith2010,Smith2013,Congreve2013} The initial and final states of singlet fission in the pentacene dimer can be described using fragment-based diabats, which are referred to as the LE and multiexcitonic (ME) states, respectively. Since the direct coupling between the LE and ME states has been shown to be small, it has been suggested that the transition from LE to ME state is mediated by a CT state that is more strongly coupled to both the LE and ME states via a superexchange mechanism.\cite{Chan2013,Berkelbach2013b,Berkelbach2014,Zeng2014} Here we investigate the performance of our $\Delta$-ALMO(MSDFT2) approach in capturing the coupling between the LE and CT diabats in the pentacene dimer as a function of intermolecular separation and monomer rotation angles. This coupling corresponds to the change in the electronic configuration illustrated in Fig.~\ref{fig:pentacene_dimer_orbs}. The orientation of the two pentacene monomers (denoted as X and Y) is that used in a recent study\cite{Zeng2014} and is based on the intermolecular configuration in crystalline pentacene. The LE state ($\ket{\mathrm{X}^\ast \mathrm{Y}}$) corresponds to the HOMO$\rightarrow$LUMO transition on monomer X while the CT state ($\ket{\mathrm{X}^{+} \mathrm{Y}^{-}}$) corresponds to the transition from the HOMO on X to the LUMO on Y.

To benchmark our results we compare to the recent results by Zeng et al.\cite{Zeng2014} who used a fourfold diabatization \cite{Nakamura2001, Nakamura2002} of adiabatic states obtained from extended multiconfigurational quasi-degenerate perturbation theory (XMCQDPT) calculations with an active space consisting of 4 electrons in 4 orbitals (4e, 4o). In particular, the two diabats we consider here are denoted as $\ket{eg}$ (excited X with ground-state Y) and $\ket{ca}$ (cationic X with anionic Y) in Ref.~\citenum{Zeng2014}. 

\begin{figure}[t!]
	\centering
	\includegraphics[width=0.45\textwidth]{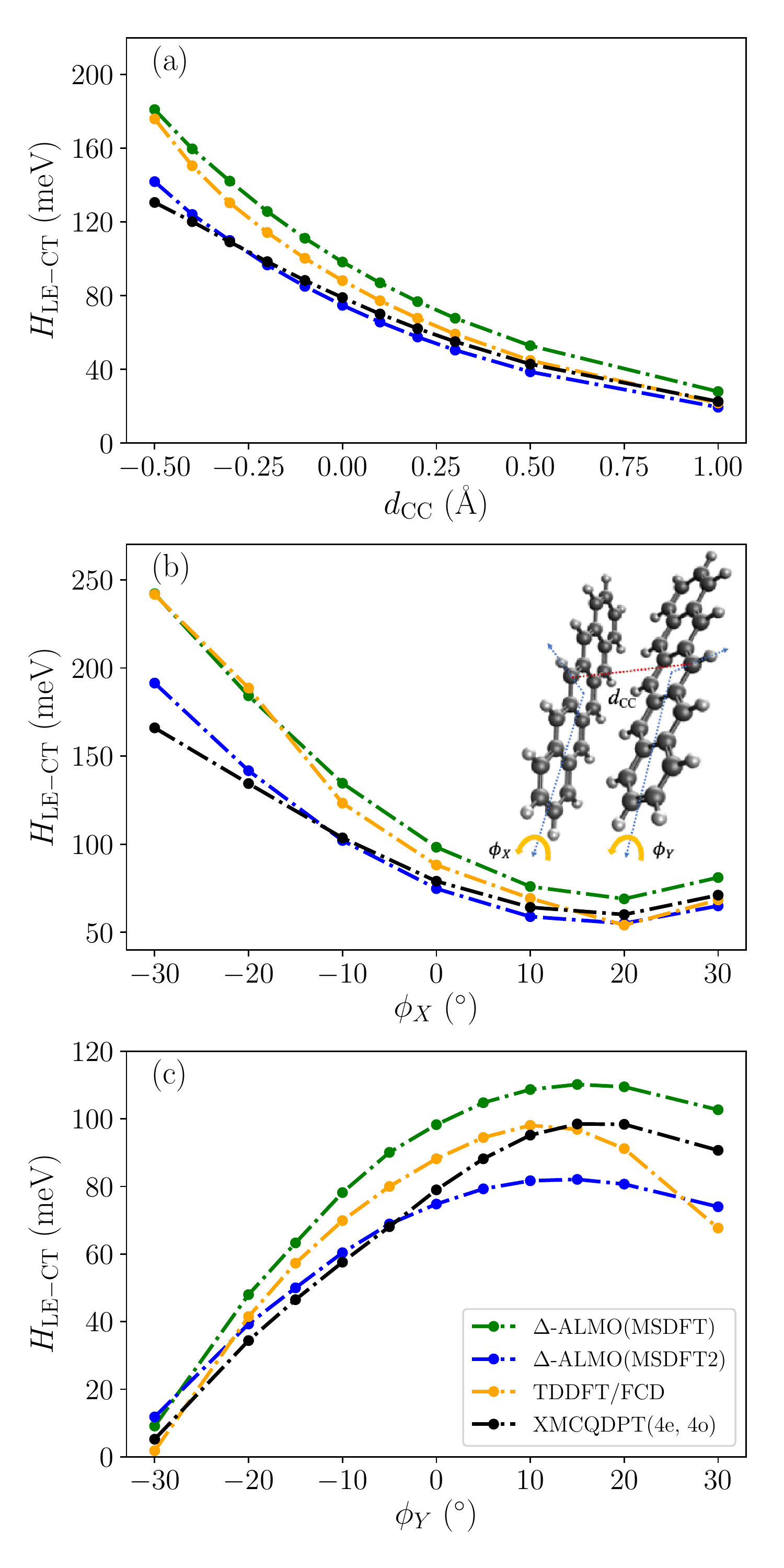}
	\caption{Distance and angular dependence of the LE-CT coupling (meV) for photoinduced ET in the pentacene dimer obtained from ALMO-$\Delta$SCF and TDDFT-based calculations. The geometric configuration of the dimer and the scanned coordinates are shown in the inset of the middle panel. (a) $H_{\text{LE-CT}}$ as a function of inter-fragment distance (measured by the top central C$\cdots$C distance, $d_{\text{CC}}$, marked in the inset); (b) and (c) $H_{\text{LE-CT}}$ as functions of the rotation angle of monomers X and Y, respectively. The LE state corresponds to the HOMO$\rightarrow$LUMO transition of monomer X and the CT state transition from the HOMO of monomer X to LUMO of monomer Y. The benchmark XMCQDPT(4e, 4o) results are from Ref.~\citenum{Zeng2014}, for which the fourfold diabatization scheme \cite{Nakamura2001, Nakamura2002} was employed to construct the diabatic Hamiltonian and extract the diabatic couplings.}
	\label{fig:pentacene_dimer_scan}
\end{figure}

Figure~\ref{fig:pentacene_dimer_scan} shows the LE-CT couplings obtained from $\Delta$-ALMO(MSDFT) and $\Delta$-ALMO(MSDFT2) as functions of the inter-monomer distance ($d_{\text{CC}}$) and the rotation angles of monomers X and Y ($\phi_X$ and $\phi_Y$) in comparison to the reference values. For the $d_{\text{CC}}$ distance and $\phi_X$ scans, shown in Figs.~\ref{fig:pentacene_dimer_scan}(a) and (b), our $\Delta$-ALMO(MSDFT2) approach gives excellent agreement with the XMCQDPT results with most of the relative errors below 10\% and the largest being 15\% at $\phi_X=-30^{\circ}$. Using our approach with the MSDFT coupling ($\Delta$-ALMO(MSDFT)), on the other hand, leads to overestimation of the LE-CT coupling at all distances and angles in these scans, with largest relative errors above 40\%. For the scan of $\phi_Y$ (Fig.~\ref{fig:pentacene_dimer_scan}(c)), $\Delta$-ALMO(MSDFT2) shows reasonably good agreement with XMCQDPT at negative angles but markedly underestimates the coupling for $\phi_Y > 0$. Excluding the data point at $\phi_Y = -30^{\circ}$ where the magnitude of the XMCQDPT result is very close to zero (5~meV), the largest (unsigned) relative error of $\Delta$-ALMO(MSDFT2) is 18\% at $\phi_Y = 30^{\circ}$. Nevertheless, $\Delta$-ALMO(MSDFT2) still qualitatively captures the change in the LE-CT coupling strength in the $\phi_Y > 0$ regime and correctly predicts the value of $\phi_Y$ (15$^\circ$) corresponding to the maximum LE-CT coupling. Again $\Delta$-ALMO(MSDFT) systematically overestimates the LE-CT coupling in the $\phi_Y$ scan, whose relative error is above 30\% when $\phi_Y < 0$ while reducing to 10-20\% when $\phi_Y > 0$. Finally, we note that at $\phi_Y = -30^{\circ}$ we invoked the MSDFT2-wc coupling scheme given in Eq.~(\ref{eq:fock_approach}) to generate the $\Delta$-ALMO(MSDFT2) result, since the condition $\vert H_{ab}^{\prime} \vert < \vert \mathcal{S}_{ab}(E_{a} + E_{b})/2 \vert$ is satisfied at this geometry. As shown in Fig.~\ref{fig:pentacene_dimer_scan}, this data point connects smoothly with the ones that were obtained from using the standard MSDFT2 coupling scheme.

We also performed FCD diabatization of adiabatic excited states obtained from TDDFT calculations. We note that the standard GMH diabatization scheme cannot be used in this case due to the lack of a uniform charge-transfer direction. For this system FCD diabatization requires the inclusion of at least the four lowest TDDFT excited states since the monomer HOMOs and LUMOs strongly couple with each other in the adiabatic picture giving rise to four frontier orbitals of the full system (from HOMO$-$1 to LUMO$+$1). The details of the 4-state FCD diabatization procedure are provided in SI Sec.~S6. As shown in Fig.~\ref{fig:pentacene_dimer_scan}, TDDFT/FCD gives poorer agreement with the reference results than $\Delta$-ALMO(MSDFT2) but in general performs similarly to $\Delta$-ALMO(MSDFT) since it also overestimates the coupling for most of scanned distances and rotation angles. In particular, TDDFT/FCD incorrectly predicts the turnover in the magnitude of coupling as a function of $\phi_Y$ giving an early maximum at $\phi_Y = 10^{\circ}$ and a too rapid decay as the angle is further increased.

\section{Conclusions}

Here, we have introduced the $\Delta$-ALMO(MSDFT2) method, which combines our ALMO-$\Delta$SCF scheme to generate excited state diabats with our MSDFT2 scheme to calculate their couplings. We have shown that our method gives excellent agreement with the diabatic couplings obtained using high-level wavefunction-based schemes for a wide variety of systems relevant to DNA damage, charge separation in donor-acceptor dyads, photoinduced electron transfer to the solvent environment, and singlet fission. In addition, our ALMO approach is more generally applicable and yields better accuracy than other DFT-based methods, such as GMH and FCD diabatization of TDDFT adiabatic states, particularly for charged systems. Unlike schemes based on the adiabatic-to-diabatic transformation, $\Delta$-ALMO(MSDFT2) gives access to variationally optimized diabatic states and thus provides easy access to nuclear gradients (i.e.~forces) necessary for quantum dynamics simulations of photoinduced processes. While having a computational cost only $\sim$2--3 times that of the corresponding ground-state DFT calculation, our $\Delta$-ALMO(MSDFT2) method yields LE-CT diabatic couplings typically within 10--20\% of benchmark values obtained from high-level wavefunction methods, such as GMH diabatization of EOM-CCSD adiabatic states. Importantly, although we used broken-symmetry $\Delta$SCF to construct the excited-state diabats, our framework is flexible and therefore compatible with constructions that, for example, explicitly account for spin symmetry adaptations. This development thus opens the door to quantum dynamics simulations of photoinduced electron and hole transfer processes using diabats constructed directly from DFT calculations in the condensed phase.

\section*{Supplementary Material}
See the supplementary material for procedures of GMH with 3 adiabatic states and FCD with a general number of adiabatic states as well as their applications to the indole-guanine complex and pentacene dimer systems; results of TDDFT and EOM-CCSD excited-state calculations required for GMH and FCD diabatization and other intermediate results; basis set sensitivity of diabatic couplings calculated with different schemes (PDF); geometries of the investigated complexes and the original data for diabatic couplings (ZIP).

\section*{Acknowledgments}
The authors greatly thank Profs.~Tao Zeng and Nandini Ananth for providing the structures and diabatic coupling data for the pentacene dimer distance and angular scans, and Prof.~Robert Cave for the guidance on the implementation of the 3-state GMH diabatization scheme. This material is based upon work supported by the National Science Foundation under Grant No.~CHE-1652960. T.E.M also acknowledges support from the Camille Dreyfus Teacher-Scholar Awards Program.  This research also used resources of the National Energy Research Scientific Computing Center (NERSC), a U.S.~Department of Energy Office of Science User Facility operated under Contract No.~DE-AC02-05CH11231.

\section*{Data Availability}
The data that supports the findings of this study are available within the article and its supplementary material.

%


\clearpage

\setcounter{section}{0}
\setcounter{equation}{0}
\setcounter{figure}{0}
\setcounter{table}{0}
\setcounter{page}{1}

\renewcommand{\theequation}{S\arabic{equation}}
\renewcommand{\thefigure}{S\arabic{figure}}
\renewcommand{\thetable}{S\arabic{table}}
\renewcommand{\thepage}{S\arabic{page}}
\renewcommand{\bibnumfmt}[1]{$^{\mathrm{S#1}}$}
\renewcommand{\citenumfont}[1]{S#1}

\title{Supporting Information for ``Excited state diabatization on the cheap using DFT: Photoinduced electron and hole transfer"}

{\maketitle}

\onecolumngrid

\section{3-state Generalized Mulliken-Hush (GMH) diabatization} \label{sec:3state_gmh}

The 3-state GMH diabatization follows the procedure described in Ref.~\citenum{Rust2002}, which we summarize here. Here, we have three adiabatic states whose energies are denoted as $E_1$, $E_2$, $E_3$ and dipole moments as $\boldsymbol{\mu}_1$, $\boldsymbol{\mu}_2$, $\boldsymbol{\mu}_3$. To generate diabatic states from these adiabats, one first needs to specify the charge-transfer direction. This can be achieved by identifying the state with significant CT character or by calculating the difference in dipole moments between each pair of adiabatic states ($\boldsymbol{\mu}_1 - \boldsymbol{\mu}_2$, $\boldsymbol{\mu}_1 - \boldsymbol{\mu}_3$, and $\boldsymbol{\mu}_2 - \boldsymbol{\mu}_3$) and finding the two that are closer to being co-linear. Without loss of generality, we denote the state of significant CT character as state 3 and the charge-transfer (CT) direction as $\hat{\mathbf{e}}_{\text{CT}} = (\hat{\mathbf{e}}_{13} + \hat{\mathbf{e}}_{23})/2$, where $\hat{\mathbf{e}}_{13} = (\boldsymbol{\mu}_{1} - \boldsymbol{\mu}_{3})/\vert \boldsymbol{\mu}_{1} - \boldsymbol{\mu}_{3} \vert$ and $\hat{\mathbf{e}}_{23} = (\boldsymbol{\mu}_{2} - \boldsymbol{\mu}_{3})/\vert \boldsymbol{\mu}_{2} - \boldsymbol{\mu}_{3} \vert$. One then constructs the projected dipole matrix ($\boldsymbol{\mu}$) as
\begin{equation}
\boldsymbol{\mu} = 
\begin{pmatrix}
\mu_{11} & \mu_{12} & \mu_{13} \\
\mu_{21} & \mu_{22} & \mu_{23} \\
\mu_{31} & \mu_{32} & \mu_{33} 
\end{pmatrix},
\end{equation}
where the diagonal elements are projections of the dipole of each state onto the CT direction, $\mu_{ii} = \boldsymbol{\mu}_i \cdot \hat{\mathbf{e}}_{\text{CT}}$ with $i \in \lbrace 1,2,3 \rbrace$, and the off-diagonal elements are projections of the transition dipole vectors between each pair of adiabatic states onto the same direction, $\mu_{ij} = \boldsymbol{\mu}_{ij} \cdot \hat{\mathbf{e}}_{\text{CT}}$ with $i, j \in \lbrace 1, 2, 3 \rbrace$ and $i \neq j$.

One then diagonalizes the projected dipole matrix, $\boldsymbol{\mu} = \mathbf{U}\mathbf{d}\mathbf{U}^{\dagger}$, and then constructs the diabatic Hamiltonian using the eigenvectors ($\mathbf{U}$) obtained:
\begin{equation}
\mathbf{H}_{\text{diab}} = \mathbf{U} 
\begin{pmatrix}
E_1 & 0 & 0 \\
0 & E_2 & 0 \\
0 & 0 & E_3
\end{pmatrix}
\mathbf{U}^{\dagger}
=
\begin{pmatrix}
H_{11} & H_{12} & H_{13} \\
H_{21} & H_{22} & H_{23} \\
H_{31} & H_{32} & H_{33} 
\end{pmatrix},
\end{equation}
which is a $3\times 3$ matrix with non-zero off-diagonal elements.  $\mathbf{H}_{\text{diab}}$, however, is not the final result since GMH requires diabats belonging to the same site remain ``adiabatic'' with respect to each other. Considering a case where the ground state (GS) and locally excited (LE) state on one site are coupled to the CT state, one needs to diagonalize a sub-block ($2\times 2$ in this case) of $\mathbf{H}_{\text{diab}}$ by performing an additional unitary transformation
\begin{equation}
\bar{\mathbf{H}}_{\text{diab}} = \mathbf{V}^{\dagger} \mathbf{H}_{\text{diab}} \mathbf{V} =
\begin{pmatrix}
E_{\text{GS}} & 0 & H_{\text{GS-CT}} \\
0 & E_{\text{LE}} & H_{\text{LE-CT}} \\
H_{\text{GS-CT}} & H_{\text{LE-CT}} & E_{\text{CT}}
\end{pmatrix},
\label{eq:gmh_same_site}
\end{equation}
where $\mathbf{V}$ has the following structure:
\begin{equation}
\mathbf{V} \equiv 
\begin{pmatrix}
V_{11} & V_{12} & 0 \\
V_{21} & V_{22} & 0 \\
0 & 0 & 1
\end{pmatrix}.
\label{eq:subblock_rotation}
\end{equation}
From the off-diagonal elements of $\bar{\mathbf{H}}_{\text{diab}}$ one can then obtain the GS-CT and LE-CT diabatic couplings.

\section{Fragment charge difference (FCD) diabatization with multiple states} \label{sec:fcd_multistate}

The FCD diabatization scheme \cite{Voityuk2002, Voityuk2013} does not require the definition of a CT direction and thus is more flexible for multi-state cases. Given a general number of adiabatic states, we denote the charge difference matrix between donor ($D$) and acceptor ($A$) fragments as $\mathbf{\Delta Q}$ , whose diagonal and off-diagonal elements can be obtained by calculating fragment charge populations using the electron density of each adiabatic state and transition density between each pair of adiabats, respectively:
\begin{equation}
\Delta Q_{ij} = Q_{ij}(D) - Q_{ij}(A) = \int_{\mathbf{r}\in D} \rho_{ij}(\mathbf{r}) \mathrm{d}\mathbf{r} - \int_{\mathbf{r}\in A} \rho_{ij}(\mathbf{r}) \mathrm{d}\mathbf{r}.
\label{eq:deltaQ}
\end{equation}
Note that although we formally introduced a partition in real-space in Eq.~(\ref{eq:deltaQ}) to define the fragment charges, in practice $\Delta Q_{ij}$ can be evaluated with any charge population schemes. In this work, we employed the simplest Mulliken population definition.\cite{Mulliken1955}

The remaining steps are similar to the procedure of GMH. One first diagonalizes $\mathbf{\Delta Q} = \mathbf{UqU}^{\dagger}$ and uses the eigenvectors to construct the diabatic Hamiltonian: $\mathbf{H}_{\text{diab}} = \mathbf{U} \boldsymbol{\varepsilon} \mathbf{U}^{\dagger}$. Here $\boldsymbol{\varepsilon}$ denotes the diagonal matrix whose (diagonal) elements are simply energies of each adiabatic state. The FCD diabatization scheme also requires the diabats on the same site to be ``adiabatic'' with respect to each other, which can be achieved by applying an additional unitary transformation in a similar manner as shown in Eqs.~(\ref{eq:gmh_same_site}) and (\ref{eq:subblock_rotation}).

\section{Indole--Guanine Complex}

\subsection{Basis set set dependence of the diabatization schemes}

\begin{figure}[h!]
	\centering
	\includegraphics[width=0.9\textwidth]{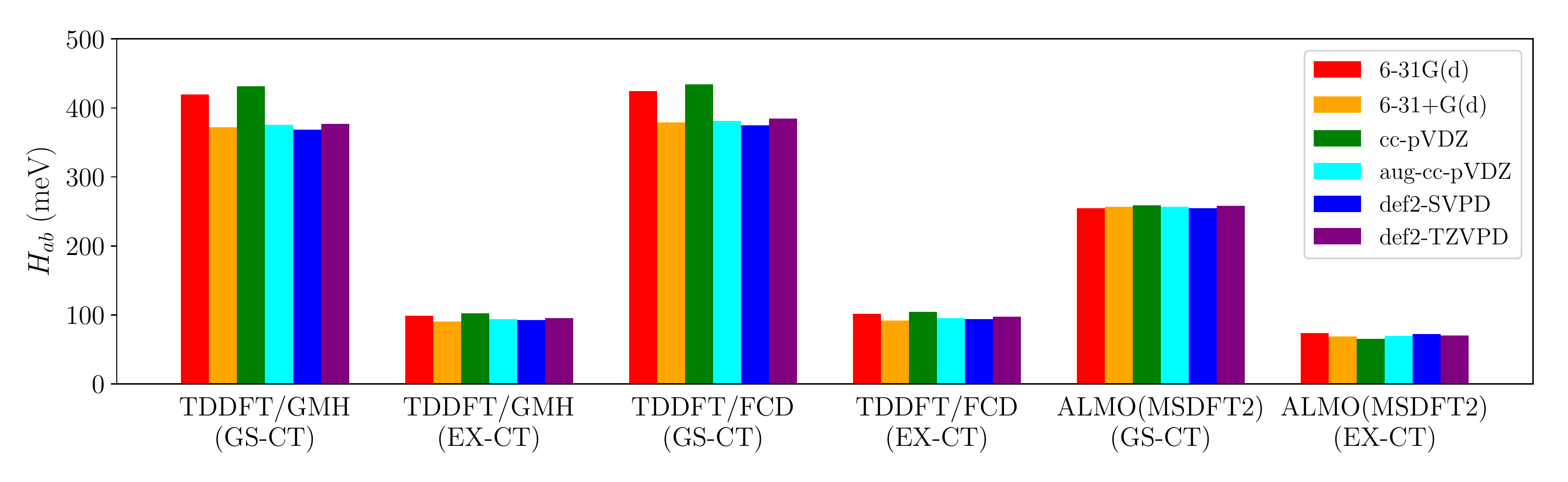}
	\caption{Performance of TDDFT/GMH, TDDFT/FCD (using 3 adiabats) and ALMO(MSDFT2) with varying basis sets in predicting the diabatic couplings (in meV) for the ground and excited state hole transfer in the cationic indole-guanine complex.}
	\label{fig:ind_guanine_basis}
\end{figure}

Figure~\ref{fig:ind_guanine_basis} shows the basis set dependence of the GS-CT and LE-CT couplings for the ground and excited state hole transfer in the cationic indole-guanine complex obtained from TDDFT/GMH, TDDFT/FCD, and ALMO(MSDFT2) calculations using the $\omega$B97X-D \cite{Chai2008} functional. The ALMO(MSDFT2) results exhibit only minimal changes across the assessed basis sets ranging from the smallest 6-31G(d)\cite{Hehre1972,Frisch1984} to the largest def2-TZVPD.\cite{Weigend2005} The TDDFT/GMH and TDDFT/FCD results are also stable with the choice of basis set in general but exhibit larger fluctuations than the ALMO(MSDFT2) results.

\subsection{3-state GMH diabatization of EOM-IP-CCSD} 

Table~\ref{tab:ind_guanine_eomip} shows the EOM-IP-CCSD/6-31+G(d) results for the first two electronic states of the cationic indole-guanine complex. It can be seen that the dipole moments of states 1 and 2 are similar while that for state 3 is clearly distinct. Therefore, we determine the charge-transfer direction using the average of $\hat{\mathbf{e}}_{13} = (\boldsymbol{\mu}_{1} - \boldsymbol{\mu}_{3})/\vert \boldsymbol{\mu}_{1} - \boldsymbol{\mu}_{3} \vert$ and $\hat{\mathbf{e}}_{23} = (\boldsymbol{\mu}_{2} - \boldsymbol{\mu}_{3})/\vert \boldsymbol{\mu}_{2} - \boldsymbol{\mu}_{3} \vert$. Following the procedure that we have described in Sec.~\ref{sec:3state_gmh}, we obtain the diabatic Hamiltonian (matrix elements in eV)
\begin{equation*}
\mathbf{H}_{\text{diab}} =
\begin{pmatrix*}[r]
7.594 & 0.102 & -0.264 \\
0.102 & 7.583 & 0.057 \\
-0.264 & 0.057 & 7.224
\end{pmatrix*}
\end{equation*}
and the adiabatic-to-diabatic transformation matrix
\begin{equation*}
\mathbf{U} =
\begin{pmatrix*}[r]
0.469 & -0.189 & 0.863 \\
0.249 & -0.909 & -0.334 \\
0.847 & 0.371 & -0.380
\end{pmatrix*}.
\end{equation*}

\begin{table}[t!]
	\centering
	\caption{EOM-IP-CCSD results for the cationic indole-guanine complex. The calculations are based on a closed-shell reference state. The excitation energies are in eV and the dipole moments are in a.u.}
	\label{tab:ind_guanine_eomip}
	\resizebox{0.35\textwidth}{!}{
		\begin{tabular}{lrrrr}
			\hline
			state & $E$ & $\mu_x$ & $\mu_y$ & $\mu_z$ \\
			\hline
			1 & 7.068 & 2.346 & -2.353 & 5.074 \\
			2 & 7.576 & 2.576 & -2.007 & 5.898 \\
			3 & 7.758 & 0.688 & -1.866 & 2.396 \\
			1--2 &  & -0.431 & 0.221 & -0.631 \\
			1--3 &  & -1.248 & 0.281 & -2.164 \\
			2--3 &  & -0.592 & -0.044 & -1.109 \\
			\hline
		\end{tabular}
	}
\end{table}

Based on the composition of each column in $\mathbf{U}$, the first diabatic state with energy 7.594 eV is dominated by the third adiabatic state and it should correspond to the CT state (Ind-G$^+$), while the second and third diabats correspond to the LE (Ind$^{+\ast}$-G) and GS (Ind$^+$-G) states. Performing an additional unitary transformation to make the LE and GS ``adiabatic'' with respect to each other yields the final diabatic Hamiltonian (in eV)
\begin{equation*}
\bar{\mathbf{H}}_{\text{diab}} =
\begin{pmatrix*}[r]
7.594 & 0.060 & 0.277 \\
0.060 & 7.592 & 0 \\
0.277 & 0 & 7.215
\end{pmatrix*}
\end{equation*}
and adiabatic-to-diabatic transformation matrix
\begin{equation*}
\bar{\mathbf{U}} =
\begin{pmatrix*}[r]
0.469 & 0.054 & -0.881 \\
0.249 & 0.950 & 0.191 \\
0.847 & -0.308 & 0.432
\end{pmatrix*}
\end{equation*}
obtained from 3-state GMH. The GS-CT and LE-CT couplings are thus 277 and 60 meV, respectively. Here we used this example to demonstrate the procedure of 3-state GMH diabatization in detail and in the discussion below we will skip the intermediate mathematical details.

\subsection{GMH and FCD diabatization of TDDFT states}

\begin{figure*}[h!]
	\centering
	\includegraphics[width=0.55\textwidth]{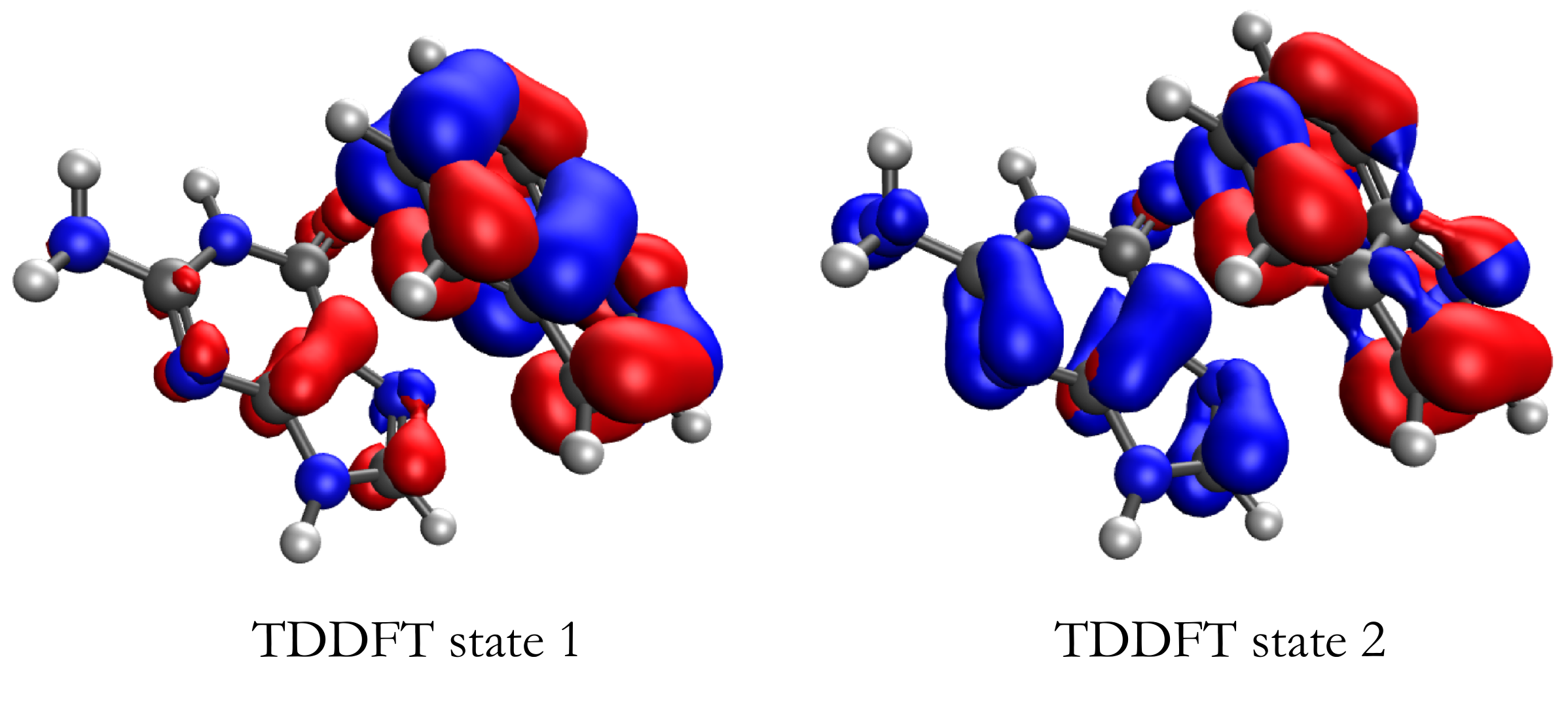}
	\caption{Detachment (blue) and attachment (red) densities of the $\beta$ spin for the first two TDDFT excited states of the cationic indole-guanine complex computed at the $\omega$B97X-D/6-31+G(d) level. The density cubes are plotted with an isovalue of 0.002 a.u. }
	\label{fig:ind_guanine_det_att}
\end{figure*}

The TDDFT calculations for the cationic indole-guanine complex were performed with an unrestricted KS-DFT reference state. When global hybrid (B3LYP, PBE0) and range-separated hybrid (CAM-B3LYP, $\omega$B97X-D, and LRC-$\omega$PBEh) functionals are employed, the relevant adiabatic states are the ground and the first two TDDFT excited states. Figure~\ref{fig:ind_guanine_det_att} shows the results of detachment-attachment density analysis \cite{HeadGordon1995} for the first two TDDFT states calculated with $\omega$B97X-D/6-31+G(d). While the first excited state (left panel) mainly involves a local excitation on the indole moiety, the second excited state exhibits marked charge-transfer character, with an electron flowing from guanine to indole. Based on this analysis, we can assign LE and CT characters to the first and second TDDFT adiabatic states, respectively. Note that when pure functionals (BLYP and PBE) are employed, the CT-dominated adiabat shifts to being the fourth TDDFT excited state.

The TDDFT results required for the 3-state GMH and FCD diabatization are shown in Table~\ref{tab:ind_guanine_tddft}. In 3-state GMH, the charge-transfer direction is determined by the average of $\mathbf{e}_1 = (\boldsymbol{\mu}_{\text{CT}} - \boldsymbol{\mu}_{\text{GS}})/ \vert \boldsymbol{\mu}_{\text{CT}} - \boldsymbol{\mu}_{\text{GS}} \vert$ and $\mathbf{e}_2 = (\boldsymbol{\mu}_{\text{CT}} - \boldsymbol{\mu}_{\text{EX}})/ \vert \boldsymbol{\mu}_{\text{CT}} - \boldsymbol{\mu}_{\text{EX}} \vert$.

\begin{table}[h!]
	\centering
	\caption{TDDFT results for the cationic indole-guanine complex with varying functionals that are required for GMH and FCD diabatization. The state with index ``0'' corresponds to the ground state. The excitation energies are in eV and the charge differences and dipole moments are in a.u.}
	\label{tab:ind_guanine_tddft}
	\resizebox{0.7\textwidth}{!}{
		\begin{tabular}{lrrrrrcrrrrr}
			\hline
			& \multicolumn{5}{c}{BLYP} && \multicolumn{5}{c}{PBE} \\
			\cline{2-6} \cline{8-12}
			state & $E$ & $\mu_x$ & $\mu_y$ & $\mu_z$ & $\Delta Q$ && $E$ & $\mu_x$ & $\mu_y$ & $\mu_z$ & $\Delta Q$ \\
			\hline
			0 & 0 & -0.366 & 0.475 & -1.602 & -0.431 && 0 & -0.322 & 0.471 & -1.586 & -0.341 \\
			1 & 0.700 & -2.534 & -1.257 & -2.245 & -1.196 && 0.712 & -2.467 & -1.264 & -2.166 & -1.092 \\
			4 & 1.511 & 1.614 & 0.746 & -1.609 & -0.046 && 1.513 & 1.557 & 0.643 & -1.685 & -0.006 \\
			0--1 &  & -0.201 & 0.055 & -0.005 & -0.035 &&  & -0.199 & 0.059 & -0.006 & -0.035 \\
			0--4 &  & 2.748 & 0.925 & 0.710 & 0.825 &&  & 2.794 & 0.935 & 0.716 & 0.841 \\
			1--4 &  & 0.008 & -0.191 & -0.074 & -0.041 &&  & 0.010 & -0.197 & -0.083 & -0.044 \\
			\hline
			& \multicolumn{5}{c}{B3LYP} && \multicolumn{5}{c}{PBE0} \\
			\cline{2-6} \cline{8-12}
			state & $E$ & $\mu_x$ & $\mu_y$ & $\mu_z$ & $\Delta Q$ && $E$ & $\mu_x$ & $\mu_y$ & $\mu_z$ & $\Delta Q$ \\
			\hline
			0 & 0 & -0.586 & 0.468 & -1.678 & -0.435 && 0 & -0.606 & 0.461 & -1.678 & -0.349 \\
			1 & 0.737 & -2.563 & -1.217 & -2.371 & -1.174 && 0.770 & -2.553 & -1.193 & -2.365 & -1.080 \\
			2 & 1.273 & 1.226 & 0.779 & -1.485 & -0.016 && 1.208 & 1.122 & 0.781 & -1.452 & 0.064 \\
			0--1 &  & -0.358 & 0.025 & -0.043 & -0.079 &&  & -0.439 & 0.008 & -0.065 & -0.104 \\
			0--2 &  & 2.796 & 0.968 & 0.811 & 0.876 &&  & 2.787 & 0.969 & 0.824 & 0.883 \\
			1--2 &  & -0.187 & -0.347 & -0.102 & -0.106 &&  & -0.277 & -0.402 & -0.135 & -0.137 \\
			\hline
			& \multicolumn{5}{c}{CAM-B3LYP} && \multicolumn{5}{c}{LRC-$\omega$PBEh} \\
			\cline{2-6} \cline{8-12}
			state & $E$ & $\mu_x$ & $\mu_y$ & $\mu_z$ & $\Delta Q$ && $E$ & $\mu_x$ & $\mu_y$ & $\mu_z$ & $\Delta Q$ \\
			\hline
			0 & 0 & -1.181 & 0.304 & -1.856 & -0.623 && 0 & -1.175 & 0.274 & -1.838 & -0.513 \\
			1 & 0.738 & -1.758 & -0.681 & -2.114 & -0.916 && 0.737 & -1.503 & -0.508 & -2.007 & -0.713 \\
			2 & 0.932 & 0.945 & 0.695 & -1.374 & -0.070 && 0.897 & 0.635 & 0.515 & -1.421 & -0.052 \\
			0--1 &  & -1.227 & -0.275 & -0.321 & -0.360 &&  & -1.410 & -0.341 & -0.383 & -0.426 \\
			0--2 &  & 2.296 & 0.873 & 0.740 & 0.752 &&  & 2.176 & 0.854 & 0.718 & 0.727 \\
			1--2 &  & -1.529 & -1.008 & -0.491 & -0.530 &&  & -1.709 & -1.093 & -0.554 & -0.592 \\
			\hline
			& \multicolumn{5}{c}{$\omega$B97X-D} && &  &  &  & \\
			\cline{2-6} 
			state & $E$ & $\mu_x$ & $\mu_y$ & $\mu_z$ & $\Delta Q$ && &  &  &  & \\
			\hline        
			0 & 0 & -1.253 & 0.270 & -1.876 & -0.612 &&  &  &  &  &  \\
			1 & 0.717 & -1.691 & -0.653 & -2.083 & -0.861 &&  &  &  &  &  \\
			2 & 0.905 & 0.948 & 0.733 & -1.327 & -0.020 &&  &  &  &  &  \\
			0--1 &  & -1.224 & -0.279 & -0.325 & -0.363 &&  &  &  &  &  \\
			0--2 &  & 2.227 & 0.866 & 0.735 & 0.738 &&  &  &  &  &  \\
			1--2 &  & -1.574 & -1.029 & -0.517 & -0.549 &&  &  &  &  &  \\
			\hline
		\end{tabular}
	}
\end{table}

\clearpage

\section{Naphthalene--TCNE complex}

Table~\ref{tab:np_tcne_intermediate} shows the intermediate coupling results that were employed to evaluate the coupling between the LE2 state (a 1:1 mix of the HOMO$\rightarrow$LUMO$+$1 and HOMO$-$1$\rightarrow$LUMO transitions on naphthalene) and the lowest-energy CT (from naphthalene's HOMO to TCNE's LUMO) state of the naphthalene--TCNE complex. Table~\ref{tab:np_tcne_ex_states} provides the results of EOM-EE-CCSD and TDDFT calculations of electronic excited states that are required in GMH and/or FCD diabatization.

\begin{table}[h!]
	\centering
	\caption{Intermediate results required for calculating the LE2-CT coupling that is relevant to the photoinduced electron transfer in the naphthalene-TCNE complex using $\Delta$-ALMO(MSDFT) and $\Delta$-ALMO(MSDFT2). The distances are in \AA\ and the diabatic couplings are in meV. ``eq'' denotes the equilibrium complex geometry obtained from Ref.~\citenum{Stein2009}.}
	\label{tab:np_tcne_intermediate}
	\resizebox{\textwidth}{!}{
		\begin{tabular}{lccccccc}
			\hline
			& \multicolumn{3}{c}{$\Delta$-ALMO(MSDFT)} && \multicolumn{3}{c}{$\Delta$-ALMO(MSDFT2)} \\
			\cline{2-4} \cline{6-8}
			distance & $H$(H$\rightarrow$L$+$1,~CT) & $H$(H$-$1$\rightarrow$L,~CT) & $H(\text{LE2-CT})$ && $H$(H$\rightarrow$L$+$1,~CT) & $H$(H$-$1$\rightarrow$L,~CT) & H(\text{LE2-CT}) \\
			\hline
			3.5 & 360.6 & 19.8 & 269.0 && 282.6 & 21.1 & 214.7 \\
			3.9 (eq) & 218.5 & 12.8 & 163.6 && 165.6 & 13.7 & 126.8 \\
			4.0 & 192.6 & 11.4 & 144.2 && 144.7 & 12.2 & 110.9 \\
			4.5 & 104.6 & 6.2 & 78.3 && 75.2 & 6.6 & 57.8 \\
			5.0 & 58.1 & 3.5 & 43.6 && 39.6 & 3.7 & 30.6 \\
			\hline
		\end{tabular}
	}
\end{table}


\begin{table}[h!]
	\centering
	\caption{EOM-EE-CCSD and TDDFT (using the $\omega$B97X-D functional) results for the first two B$_1$ excited states of the naphthalene-TCNE complex at varying intermolecular distances required for GMH and FCD diabatization. The energies are in eV and the charge differences and dipole moments are in a.u. }
	\label{tab:np_tcne_ex_states}
	\resizebox{0.6\textwidth}{!}{
		\begin{tabular}{llcccccccccc}
			\hline
			&  & \multicolumn{4}{c}{EOM-EE-CCSD} && \multicolumn{5}{c}{TDDFT~($\omega$B97X-D)} \\
			\cline{3-6} \cline{8-12}
			&  & $E$ & $\mu_x$ & $\mu_y$ & $\mu_z$ && $E$ & $\mu_x$ & $\mu_y$ & $\mu_z$ & $\Delta Q$ \\
			\hline
			3.9 \AA & 1 & 2.958 & 0 & 0 & -6.809 && 1.969 & 0 & 0 & 7.591 & 2.077 \\
			(eq) & 2 & 4.426 & 0 & 0 & -0.346 && 4.665 & 0 & 0 & 0.325 & 0.089 \\
			& 1--2 &  & 0 & 0 & -0.574 &&  & 0 & 0 & -0.309 & -0.093 \\
			\hline
			3.5 \AA & 1 & 2.790 & 0 & 0 & -5.915 && 1.863 & 0 & 0 & 6.833 & 2.122 \\
			& 2 & 4.430 & 0 & 0 & -0.437 && 4.661 & 0 & 0 & 0.391 & 0.160 \\
			& 1--2 &  & 0 & 0 & -0.753 &&  & 0 & 0 & -0.459 & -0.147 \\
			\hline
			4.0 \AA & 1 & 2.995 & 0 & 0 & -7.028 && 1.993 & 0 & 0 & 7.778 & 2.069 \\
			& 2 & 4.424 & 0 & 0 & -0.328 && 4.665 & 0 & 0 & 0.309 & 0.077 \\
			& 1--2 &  & 0 & 0 & -0.532 &&  & 0 & 0 & -0.277 & -0.083 \\
			\hline
			4.5 \AA & 1 & 3.153 & 0 & 0 & -8.076 && 2.117 & 0 & 0 & 8.694 & 2.047 \\
			& 2 & 4.420 & 0 & 0 & -0.255 && 4.664 & 0 & 0 & 0.246 & 0.045 \\
			& 1--2 &  & 0 & 0 & -0.358 &&  & 0 & 0 & -0.161 & -0.045 \\
			\hline
			5.0 \AA & 1 & 3.289 & 0 & 0 & -9.089 && 2.238 & 0 & 0 & 9.607 & 2.028 \\
			& 2 & 4.417 & 0 & 0 & -0.205 && 4.662 & 0 & 0 & 0.201 & 0.024 \\
			& 1--2 &  & 0 & 0 & -0.233 &&  & 0 & 0 & -0.094 & -0.024 \\
			\hline
		\end{tabular}
	}
\end{table}


\section{Naphthol--Chloroform complex}

Table~\ref{tab:npoh_chcl3_Eex} shows the comparison between the energies of the LE1, LE2, and CT states of the 1-naphthol--CHCl$_3$ (NpOH--CHCl$_3$) complex obtained from TDDFT, ALMO-$\Delta$SCF, and EOM-EE-CCSD calculations. The LE1 state corresponds to NpOH's HOMO$\rightarrow$LUMO transition and the LE2 state is dominated by NpOH's HOMO$\rightarrow$LUMO+1 transition. The CT state corresponds to the charge transfer from NpOH's HOMO to CHCl$_3$'s LUMO. Table~\ref{tab:npoh_chcl3_gmh_fcd} provides the results of EOM-EE-CCSD and TDDFT calculations of electronic excited states that are required in GMH and/or FCD diabatization.

\begin{table}[h!]
	\centering
	\caption{Excitation energies (in eV) for the LE1, LE2, and CT states of the 1-naphthol--CHCl$_3$ complex calculated using TDDFT, ALMO-$\Delta$SCF, and EOM-EE-CCSD. The first two methods employ the $\omega$B97X-D functional and all calculations are performed with the 6-31+G(d) basis.}
	\label{tab:npoh_chcl3_Eex}
	\resizebox{0.33\textwidth}{!}{
		\begin{tabular}{lrrr}
			\hline
			& LE1 & LE2 & CT \\
			\hline
			TDDFT & 4.75 & 4.59 & 5.56 \\
			ALMO-$\Delta$SCF & 3.96 & 4.52 & 5.79 \\
			EOM-EE-CCSD & 5.00 & 4.37 & 6.27 \\
			\hline
		\end{tabular}
	}
\end{table}


\begin{table}[h!]
	\centering
	\caption{EOM-EE-CCSD and TDDFT (using the $\omega$B97X-D functional) results for the excited states that are relevant to photoinduced electron transfer in the 1-naphthol-CHCl$_3$ complex and required for GMH and FCD diabatization. The energies are in eV and the charge differences and dipole moments are in a.u. }
	\label{tab:npoh_chcl3_gmh_fcd}
	\resizebox{0.66\textwidth}{!}{
		\begin{tabular}{lcccccccccc}
			\hline
			& \multicolumn{4}{c}{EOM-EE-CCSD} && \multicolumn{5}{c}{TDDFT($\omega$B97X-D)} \\
			\cline{2-5} \cline{7-11}
			& $E$ & $\mu_x$ & $\mu_y$ & $\mu_z$ && $E$ & $\mu_x$ & $\mu_y$ & $\mu_z$ & $\Delta Q$ \\
			\hline
			LE1 & 4.999 & -0.034 & -1.782 & -0.574 && 4.746 & -0.608 & -0.532 & -0.406 & 0.023 \\
			LE2 & 4.369 & 0.041 & -1.105 & -0.692 && 4.589 & -0.910 & -1.395 & -0.180 & 0.003 \\
			CT & 6.274 & -1.052 & -1.303 & -3.145 && 5.558 & -3.991 & 0.372 & -6.271 & 1.804 \\
			LE1-CT &  & 0.545 & 0.717 & -0.257 &&  & -0.267 & 0.071 & -0.440 & -0.167 \\
			LE2-CT &  & 0.057 & 0.008 & -0.053 &&  & 0.035 & -0.022 & 0.000 & 0.023 \\
			\hline
		\end{tabular}
	}
\end{table}

\section{Pentacene dimer}

Here we use the TDDFT results for the pentacene dimer at the ``reference'' structure to demonstrate in detail how we obtained the LE-CT coupling for this system using 4-state FCD. Table~S7 shows the TDDFT excitation energies and donor-acceptor charge differences evaluated using the electron density of each excited state as well as the transition density between each pair of states. The full set of TDDFT results for the pentacene dimer at all scanned distances and angles is available in the spreadsheet that is provided in the Supplementary Material. According to the $\Delta Q$ value of each state, states 1 and 3 exhibit partial CT character from the acceptor ($Y$) to the donor ($X$), state 4 is dominated by CT from $X$ to $Y$, while state 2, on the other hand, exhibits almost no CT character. 

\begin{table}[h!]
	\centering
	\caption{TDDFT results (with $\omega$B97X-D) for the lowest four excited states of the pentacene dimer at the ``reference'' geometry. The energies are in eV and the charge differences are in a.u.}
	\label{tab:pentacene_dimer}
	\resizebox{0.66\textwidth}{!}{
		\begin{tabular}{lrrrrrrrrrr}
			\hline
			& 1 & 2 & 3 & 4 & 1--2 & 1--3 & 1--4 & 2--3 & 2--4 & 3--4 \\
			\hline
			$E$ & 2.516 & 2.656 & 2.745 & 3.415 &  &  &  &  &  &  \\
			$\Delta Q$ & -1.303 & 0.001 & -0.620 & 1.886 & 0.021 & 0.958 & 0.111 & -0.029 & 0.003 & 0.301 \\
			\hline
		\end{tabular}
	}
\end{table}

The procedure of FCD diabatization for a general number of states has been provided in Sec.~\ref{sec:fcd_multistate}. Using the data in Table~\ref{tab:pentacene_dimer}, we have
\begin{equation*}
\mathbf{\Delta Q} = 
\begin{pmatrix*}[r]
-1.303 & 0.021 & 0.958 & 0.111 \\
0.021 & 0.001 & -0.029 & 0.003 \\
0.958 & -0.029 & -0.620 & 0.301 \\
0.111 & 0.003 & 0.301 & 1.886 \\
\end{pmatrix*}
\end{equation*}
and
\begin{equation*}
\boldsymbol{\varepsilon} = 
\begin{pmatrix*}[r]
2.516 & 0 & 0 & 0 \\
0 & 2.656 & 0 & 0 \\
0 & 0 & 2.745 & 0 \\
0 & 0 & 0 & 3.415
\end{pmatrix*}.
\end{equation*}
Diagonalizing $\mathbf{\Delta Q}$ and using its eigenvectors,
\begin{equation*}
\mathbf{U} = 
\begin{pmatrix*}[r]
-0.815 & 0.388 & 0.424 & 0.076 \\
0.017 & 0.754 & -0.657 & 0.000 \\
0.579 & 0.519 & 0.612 & 0.145 \\
-0.022 & -0.106 & -0.122 & 0.987 \\
\end{pmatrix*},
\end{equation*}
to construct $\mathbf{H}_{\text{diab}} = \mathbf{U} \boldsymbol{\varepsilon} \mathbf{U}^{\dagger}$ yields
\begin{equation*}
\mathbf{H}_{\text{diab}} = 
\begin{pmatrix*}[r]
2.594 & 0.073 & 0.082 & 0.000 \\
0.073 & 2.668 & 0.015 & -0.077 \\
0.082 & 0.015 & 2.676 & -0.088 \\
0.000 & -0.077 & -0.088 & 3.395 \\
\end{pmatrix*}
\end{equation*}
Since in this case there are no two states that reside on the same site, no additional unitary transformation is needed and $\mathbf{H}_{\text{diab}}$ is the final result for the diabatic Hamiltonian.

Using the eigenvalues of $\mathbf{\Delta Q}$, $\mathbf{q} = [-1.981, -0.009, 0.015, 1.939]$, one can identify the characters of the generated diabatic states: the first and last states correspond to the $Y\rightarrow X$ and $X \rightarrow Y$ charge transfer, respectively, and the middle two states are of an LE character. Analysis of their corresponding vectors in the adiabatic-to-diabatic transformation matrix ($\mathbf{U}$) and the amplitudes of the TDDFT adiabatic states reveals that the third state here corresponds to the LE state on monomer $X$. Hence, the diabatic coupling between LE($X$) and CT($X\rightarrow Y$) that we consider in this work corresponds to the off-diagonal element at the 3rd row and 4th column, which is 88 meV.

The same procedure was applied to all the scanned distances and angles to generate the TDDFT/FCD results shown in Fig.~10 in the main paper, except for the couplings at $d_{\text{CC}} = -0.5$~\AA\ or $\phi_X = 20^{\circ}$ and 30$^{\circ}$. With the $d_{\text{CC}} = -0.5$ \AA\ structure, the 4th TDDFT excited state strongly mixes with the 5th one such that in total 5 adiabatic states need to be included in the FCD diabatization, and when $\phi_X = 20^{\circ}$ and 30$^{\circ}$ the LE state on monomer $X$ corresponds to the 2nd (rather than the 3rd) diabatic state obtained from diagonalization of the $\mathbf{\Delta Q}$ matrix.

\end{document}